 \documentclass[manuscript]{aastex}

\usepackage[usenames,dvips]{color}

\usepackage[usenames,dvips]{color}

\slugcomment{Not to appear in Nonlearned J., 45.}

\shorttitle{Structure and Spectra from Disk and Corona in High-Luminosity AGNs}
\shortauthors{Liu et al.}

\def\mnras{MNRAS}
\def\apj{ApJ}
\def\apjl{ApJL}

\def\apjs{ApJS}
\def\aap{A\&A}

\def\pasj{PASJ}

\begin{document}


\title{The Structure and Spectral Features of a Thin Disk and Evaporation-Fed Corona in High-Luminosity AGNs}


\author{J.Y. Liu}
\affil{National Astronomical Observatories / Yunnan Observatory, Chinese Academy of Sciences,
Kunming 650011, China}
\affil{Key Laboratory for the Structure and Evolution of Celestial Objects, Chinese Academy of Sciences, Kunming 650011, China}
\email{ljy0807@ynao.ac.cn}
\and

\author{B. F. Liu and E. L. Qiao }
\affil{National Astronomical Observatories, Chinese Academy of
Sciences, Beijing 100012, China}
\email{bfliu@nao.cas.cn}
\and
\author{S. Mineshige}
\affil{Department of Astronomy, Graduate School of Science, Kyoto University, Kyoto 606-8502, Japan}

\begin{abstract}
We investigate the accretion process in high-luminosity AGNs (HLAGNs) in
the scenario of the disk evaporation
model. Based on this model, the thin disk can
extend down to the innermost stable circular orbit (ISCO) at accretion rates higher
than $0.02\dot{M}_{\rm Edd}$; while the corona is weak since part of the coronal gas is cooled by strong inverse Compton scattering of the disk photons. This implies that
the corona cannot produce as strong X-ray radiation as observed in HLAGNs
with large Eddington ratio. In addition to the viscous heating, other heating
to the corona is necessary to interpret HLAGN. In this paper, we assume that a part of
accretion energy released in the disk is transported into the corona,
heating up the electrons and thereby radiated away.
We for the first time, compute the corona structure with additional heating, taking fully into account the mass supply to the corona and find that the corona could indeed survive at
higher accretion rates and its radiation power increases. The spectra composed
of bremsstrahlung and Compton radiation are also calculated. Our calculations
show that the Compton dominated spectrum becomes harder with the increase
of energy fraction ($f$) liberating in the corona, and the photon index for hard X-ray($2-10~\rm keV$)
is $2.2 < \Gamma < 2.7 $.  We discuss possible heating mechanisms for the corona.
Combining the energy fraction transported to the corona
with the accretion rate by magnetic heating, we find that the hard X-ray
spectrum becomes steeper at larger accretion rate and the bolometric correction factor ($L_{\rm bol}/L_{\rm 2-10keV}$) increases with increasing accretion rate for $f<\frac{8}{35}$, which is roughly consistent with the observational results.

\end{abstract}



\section{INTRODUCTION}

   Accretion of gas onto the central supermassive black hole is
one of the fundamental astrophysical processes responsible for
high-efficient energy release in active galactic nuclei (AGNs).
According to the difference in the luminosity, AGNs are classified
into two types, low-luminosity AGNs (LLAGNs) and high-luminosity AGNs (HLAGNs).
The LLAGNs show hard power-law X-rays and/or radio jets; while the HLAGNs are characterized by the big
blue bump around UV waveband, soft X-ray excess, Fe $K_{\alpha}$ lines
at about 6.4 keV, the power-law
X-rays with photon index $\Gamma \sim 1.9$ and cutoff at about a few of $100~\rm keV$.
It is speculated that the different characteristics of LLAGNs and HLAGNs
are caused by  different accretion mechanism at different accretion rates,
 in a similar way to black hole X-ray binaries. The accretion in LLAGNs is
  via an advection dominated accretion flow (ADAF) (Ichimaru
1977; Narayan \& Yi 1994, 1995a, 1995b), in which most of the accretion energy is
stored in the gas as entropy and only a small fraction is radiated in hard X-rays(see Kato, Fukue \& Mineshige 2008 for a review).  While in HLAGNs,
the common accepted accretion model is a hybrid
accretion disk, i.e., a cool disk is embedded in a hot
corona. The optical to UV emission is from the optically thick and
geometrically thin accretion disk (e.g., Shakura \& Sunyaev 1973;
Lynden-Bell \& Pringle 1974) and the hard X-ray is from Compton scattering of the soft photons from the disk by the hot electrons in the optically thin corona with $T_{\rm e}\sim 10^9\rm~K$.

Observations show that the X-ray luminosity contributes a large fraction to the bolometric luminosity in HLAGNs. This can be seen from individual spectra (e.g. Elvis et al.1994; Ho 2008). The hard X-ray bolometric correction, which is defined as the ratio of bolometric luminosity to $2-10~\rm keV$ luminosity, is typically around $40-70$ with Eddington ratio above 0.1, while below 0.1 it is typically $15-25$ (Vasudevan \& Fabian 2009), which confirms the importance of X-ray radiation in LLAGNs.  Fits to the X-ray spectra of HLAGNs by Comptonization model yield an electron temperature of $\sim 10^9$K  in the optically thin hot corona (e.g. Liu et al. 2003).
A question arises as, how can the gas in the corona retain at so high a temperature when continuously radiates strong X-rays in HLAGNs? Where is the heating energy from?

Theoretically, gas accreted  from interstellar medium or secondary is cold. In the way of accretion to the central black hole, some of the gas evaporates to the corona and then accretes inward in the corona.
In HLAGNs, gas is dominantly accreted via a thin disk, only a small fraction of gas is accreted in the hot corona. Therefore, the energy released in the corona through accretion is much smaller than that in the disk. This implies that the radiation from the hot corona is much weaker than that from the disk in a steady disk-corona flow. With strong radiation in X-rays as observed in HLAGNs, the corona is over-cooled in a thermal timescale and collapses to the disk unless there is other energy supply to the corona besides the viscous heating.

Recent investigations on the spectral energy distribution (SED) in HLAGNs found that the bolometric correction increases with  increasing Eddington
 ratio $L_{\rm bol}/L_{\rm Edd}$
 ($L_{\rm Edd}=1.26\times10^{46}(M_{\rm BH}/10^8M_{\odot})\rm {ergs~s^{-1}}$), whereas independent on the black hole mass (Vasudevan \& Fabian 2007, 2009; Lu \& Yu 1999; Wang et al. 2004; Liu et al. 2009b). This correlation provides clues and constraints to possible heating mechanisms to the disk corona in HLAGNs.

 In this work, we study the corona flow at high accretion rates in the frame of disk corona evaporation/condensation model. By inclusion of additional heating to the corona, we investigate the detailed interaction, i.e. the mass and energy exchange between  disk and corona, we self-consistently determine the corona structure and disk radiation features. The spectrum from the disk and corona is calculated by Monte Carlo simulations. Comparing the spectrum with observational features, we discuss the possible heating mechanism in HLAGNs. Our aim is to interpret the SED and its correlation with Eddington ratio.

 HLAGNs include  radio-loud and radio-quiet Seyfert 1s and QSOs. Since
the radio-loud AGNs are assumed to be powered by the jet and the
beaming effect will dilute the X-ray radiation from the central
accretion engine, here we focus on accretion mechanism for the radio-quiet AGNs.
Most narrow line Seyfert 1s are also radio-quiet HLAGNs. However,
since they accrete at rate near Eddington accretion rate and the character of their spectra (e.g. soft X-ray excess and steeper/softer hard X-ray emission) are dramatically different from other HLAGNs, they are not include here and will be investigated in the future work.

 The paper is organized as follows. In section 2 we present the model.
  The numerical results are shown in the section 3.
Section 4 and Section 5 are the discussion and conclusion, respectively.

\section{MODEL}
\subsection{Conception of the Model}
 The disk evaporation model was proposed by Meyer \& Meyer-Hofmeister (1994) for dwarf novae and developed to black hole binaries by Meyer, Liu \& Meyer-Hofmeister (2000a). In this model, an optically thin, hot corona is assumed to lie above a thin disk.  Vertical conduction is important since the disk temperature is much lower than that of the corona. As the corona radiates inefficiently, the conductive heat accumulates downwards and at the transition layer it heats up some cool gas until an equilibrium density is reached so as to radiate completely the conductive flux. The evaporated gas is then accreted through the corona to the central black hole.  Such a corona is heated up by viscosity and the accretion is supplied by mass evaporation from the disk. Further investigations on effects of decoupling between ions and electrons and Compton cooling in the corona (Liu et al. 2002a), the magnetic pressure (Meyer-Hofmeister \& Meyer 2001; Qian, Liu \& Wu 2007), the change of viscosity parameter (Qiao \& Liu 2009) and the possible condensation are included.  The model has been successfully used to interpret the state transition and observational features in low/hard state of the black hole binaries.  As the disk corona is independent on the black hole mass, it is also applied to AGNs in explaining different types of spectra and the absence of the broad line region in LLAGNs (Liu et al. 2009a). Nevertheless, when it is compared with observations in HLAGN, we find that in the corona strong Compton cooling leads to the corona so weak
 that it cannot produce the hard X-ray as high as observed. There must be other energy sources besides the viscous heating for the corona.

 There are some pioneer works assuming that a certain
 fraction of local gravitational energy is liberated directly in
 the corona (e.g. Nakamura \& Osaki 1993; Haardt \& Maraschi 1991, 1993),  particularly in the frame of a magnetized disk corona (e.g., Di Matteo 1998; Di Matteo, Celotti, \& Fabian 1999; Miller \& Stone 2000; Merloni \& Fabian 2002; Liu et al. 2002b, 2003; Kawanaka, Kato, \& Mineshige 2008; Cao 2009).  In these investigations it is demonstrated that the corona can indeed produce as strong X-ray as observed  in HLAGNs if a large fraction of accretion energy is released in the corona.  Nevertheless, the mass exchange between the two-phase accretion flows was not taken into account.  Here we consider both radiation and mass coupling between disk and corona and study the detailed vertical structure of the coronal flow above a thin disk supposing some fraction ($f$) of the gravitational energy is added into the corona.

\subsection{Physical Processes and Differential Equations}
The corona can be described by the following equations.
Equation of state
\begin{equation}\label{e:pressure}
 \centering
 P={\Re \rho \over 2\mu}(T_{\rm i}+T_{\rm e}),
\end{equation}
where $\mu=0.62$ is the molecular weight assuming a standard
chemical composition ($X=0.75, Y=0.25$) for the corona. Here we
assume $n_{\rm i}=n_{\rm e}$, which is strictly true only for a pure
hydrogen plasma.

Equation of continuity
\begin{equation}\label{e:continue}
\centering
 {d\over dz}(\rho v_{\rm z})=\eta_{\rm M}{2\over R}\rho v_{\rm R} -{2z\over
R^2+z^2}\rho v_{\rm z},
\end{equation}
where $v_{\rm R}=-\alpha {V_{\rm s}^2 \over \Omega R}$ is the
radial component of velocity, $V_{\rm s}=({P \over \rho})^{1/2}$ is
the isothermal sound speed, $v_{\rm \varphi}=\sqrt{G M \over R}
(1+{z^2 \over R^2})^{-4/3}$ is the angular component of velocity.
Here, the partial derivative of mass flux with respect to radius  is approximated as, ${1\over R}{\partial R\rho v_R \over \partial R }\approx-\eta_{\rm M}{2\over R}\rho v_{\rm R}$, where  $\eta_{\rm M}$ depends on the net mass gain/loss rate through the radial boundaries (Meyer-Hofmeister \& Meyer 2003; Liu et al. 2004). In this work, $\eta_{\rm M}=1$ is taken for the case that no mass enters through the outer boundary and all the mass flowing in the corona is contributed by the evaporation (Meyer et al. 2000a).

Equation of the $z$-component of momentum is
\begin{equation}\label{e:mdot}
\rho v_{\rm z} {dv_{\rm z}\over dz}=-{dP\over dz}-\rho {GMz\over
(R^2+z^2)^{3/2}}.
\end{equation}

In the hot corona, because of heavier mass of ions than that of
electrons, the viscous heating raises the ion temperature first and
ions are cooling by coulomb collision with electrons and both radial and vertical advection. Here we don't include direct heat to electrons.  The ion thermal conduction is not taken into account because of the  long mean free path compared to the vertical scale height of the corona.  So the energy equation for ions is
\begin{equation}\label{e:ions}
\begin{array}{l}
{d\over dz}\left\{\rho_{\rm i} v_{\rm z} \left[{v^2\over 2}+{\gamma\over
\gamma-1}{P_{\rm i}\over \rho_{\rm i}}-{GM\over (R^2+z^2)^{1\over
2}}\right]\right\}\\
={3\over 2}\alpha P\Omega-q_{\rm ie}\\
+{\eta_{\rm E}}{2\over R}\rho_{\rm i} v_{\rm R}
\left[{v^2\over 2}+{\gamma\over \gamma-1}{P_{\rm i}\over \rho_{\rm i}}-{GM\over (R^2+z^2)^{1\over 2}}\right]\\
-{2z\over {R^2+z^2}}\left\{\rho_{\rm i} v_{\rm z} \left[{v^2\over
2}+{\gamma\over \gamma-1}{P_{\rm i}\over \rho_{\rm i}}-{GM\over
(R^2+z^2)^{1\over 2}}\right]\right\},
\end{array}
\end{equation}
where $\eta_{\rm E}$ is the energy modification parameter, $\eta_{\rm E}=\eta_{\rm M}+0.5$. The difference between $\eta_{\rm E}$ and $\eta_{\rm M}$ comes from the derivative of energy flux with respect to radius,  ${1\over R}{\partial R\rho v_{\rm R} \epsilon\over \partial R }\approx-\eta_M{2\over R}\rho v_{\rm R}  \epsilon + \rho v_{\rm R} {\partial \epsilon\over \partial R}=-(\eta_M+0.5){2\over R}\rho v_{\rm R}  \epsilon$,
where $\epsilon\equiv{v^2\over 2}+{\gamma\over \gamma-1}{P_{\rm i}\over \rho_{\rm i}}-{GM\over (R^2+z^2)^{1\over 2}}$ and its derivative
${\partial \epsilon\over \partial R}\approx -{\epsilon \over R}$
 since the potential, kinetic, and thermal specific energies all scale approximately as $1/R$ (Meyer-Hofmeister \& Meyer 2003; Liu et al. 2004).
In equation (\ref{e:ions}), $q_{\rm ie}$ is the exchange rate of energy between
electrons and ions through coulomb collision and is described as
\begin{equation}
{q_{\rm ie}}={\bigg({2\over \pi}\bigg)}^{1\over 2}{3\over 2}{m_{\rm e}\over
m_{\rm i}}{\ln\Lambda}{\sigma_{\rm T} c n_{\rm e} n_{\rm i}}(\kappa T_{\rm i}-\kappa T_{\rm e})
{{1+{T_{\rm *}}^{1\over 2}}\over {{T_{\rm *}}^{3\over 2}}},
\end{equation}

in which
\begin{equation}
T_{\rm *}={{\kappa T_{\rm e}}\over{m_{\rm e} c^2}}\bigg(1+{m_{\rm e}\over m_{\rm i}}{T_{\rm i}\over
T_{\rm e}}\bigg),
\end{equation}
where $m_{\rm i}$ and $m_{\rm e}$ are the proton and electron masses, $\kappa$ is the
Boltzmann constant, $c$ is the light speed, $\sigma_{\rm T}$ is the Thomson
scattering cross section and $\ln\Lambda=20$ is the Coulomb logarithm.

Different from the previous works, besides the viscous heating, a fraction of
local gravitational  energy is assumed to directly heat the electrons in corona, i.e.,
\begin{equation}\label{Qadd}
  Q_{\rm
add}=\frac{3GM\dot M}{8\pi R^3}\left[1-(3R_{\rm
s}/R)^{1/2}\right]\times f,
\end{equation}
 where $\dot{M}$ is the total accretion rate in the disk corona. The physical meaning of this additional heating term will be discussed in the section of Discussion.  To simplify the calculations, this additional heating flux is supposed to distribute along the vertical direction in the form similar to  that of Compton cooling, i.e.
\begin{equation}
q_{\rm add}={\frac{4\kappa T_{\rm e}}{m_{\rm e} c^2}}n_{\rm e}
\sigma_{\rm T} c {\frac{a T_{\rm eff0} ^4}{2}},
\end{equation}
where
\begin{equation}
T_{\rm eff0}=\left\{\frac{3GM\dot M}{8\pi R^3 \sigma}\left[1-(3R_{\rm
s}/R)^{1/2}\right]\times \lambda\right\}^{\frac{1}{4}}
\end{equation}
and $\lambda  $ is determined by the integration along $z$ from the lower boundary to upper boundary, $\int_{z_0}^{z_1}q_{\rm add}dz=Q_{\rm add}$, which gives the relation of $f$ and $\lambda$,
\begin{equation}\label{equa-f}
f=2\lambda \int_{z_0}^{z_1}{\frac{4\kappa T_{\rm e}}{m_{\rm e} c^2}}n_{\rm e}
\sigma_{\rm T}dz.
\end{equation}

Therefore,  the energy equation for both the ions and electrons is
\begin{equation}\label{e:total energy}
\begin{array}{l}
{\frac{d}{dz}\left\{\rho {v}_{\rm z}\left[{v^2\over
2}+{\gamma\over\gamma-1}{P\over\rho}
-{GM\over\left(R^2+z^2\right)^{1/2}}\right]
 + F_{\rm c} \right\}}\\
=\frac{3}{2}\alpha P{\mit\Omega}+q_{\rm add}-n_{\rm e}n_{\rm i}L(T_e)-q_{\rm Cmp}\\
+\eta_{\rm E}{2\over R}\rho v_{\rm R} \left[{v^2\over
2}+{\gamma\over\gamma-1}{P\over\rho}
-{GM\over\left(R^2+z^2\right]^{1/2}}\right]\\
-{2z\over R^2+z^2}\left\{\rho v_{\rm z}\left[{v^2\over
2}+{\gamma\over\gamma-1}{P\over\rho}-
{GM\over\left(R^2+z^2\right)^{1/2}}\right] +F_{\rm c}\right\}.
\end{array}
\end{equation}
In this equation, $n_{\rm e}n_{\rm i}L(T_e)$ is the bremsstrahlung
cooling rate and $F_{\rm c}$ is the thermal conduction (Spitzer
1962),
\begin{equation}\label{e:fc}
F_{\rm c}=-\kappa_{\rm 0}T_{\rm e}^{5\over2}{dT_{\rm e}\over dz},
\end{equation}
with $\kappa_0 = 10^{-6}{\rm erg\,s^{-1}cm^{-1}K^{-{7\over2}}}$ for fully
ionized plasma.

 For compton cooling rate, $q_{\rm Cmp}$, we mainly consider inverse Compton scattering of the soft photons from the disk, while the ones contributed from the bremsstrahlung cooling is not included.
\begin{equation}
q_{\rm Cmp} = {\frac{4\kappa T_{\rm e}}{m_{\rm e} c^2}}n_{\rm e} \sigma_{\rm T} c {\frac{a
T_{\rm eff} ^4}{2}},
\end{equation}
where $a$ the radiation constant. $T_{\rm eff}$ is the effective temperature of the underlying thin disk, fulfilling
\begin{equation}\label{e:teff}
\sigma T_{\rm eff}^4=\frac{3GM(\dot M_{\rm d}-f\dot M)}{8\pi R^3 }\left[1-(3R_{\rm
s}/R)^{1/2}\right],
\end{equation}
where $\dot M_{\rm d}$ is the mass accretion rate in the thin
disk, which depends on the distance because of evaporation,
\begin{equation}
\dot M_{\rm d} (R)= \dot{M}-\dot M_{\rm evap} (R),
\end{equation}
with $\dot M_{\rm evap}(R)$  the integrated
evaporation rate  from the outer edge of the disk to the distance $R$,  which can be approximated by the one-zone evaporation rate, $\dot M_{\rm evap}=2\pi R^2(\rho v_{\rm z})_{0}$ ( see Liu et al. 2002a).
 The negative term associated with $-f\dot M$ in Eq.(\ref{e:teff}) corresponds to the transportation of accretion energy from the disk to the corona, which results in a lower temperature in the disk. Here, we have not considered back reaction (heating of a thin disk by coronal illumination) and its justification will be discussed later.

These five differential equations,
Eqs. (\ref{e:continue}), (\ref{e:mdot}), (\ref{e:ions}),
(\ref{e:total energy}), (\ref{e:fc}), which contain five variables $P(z)$,
$T_{\rm i}(z)$, $T_{\rm e}(z)$, $F_{\rm c}(z)$, and $\rho v_{\rm z}$, can be solved with five boundary conditions.
\subsection{Boundary Conditions and Numerical Method of Computation}
At the lower boundary, the
temperature of the gas should be the effective temperature of the
accretion disk. Liu et al. (1995) showed that the coronal temperature increases
from effective temperature to $10^{6.5}$ K in a very thin layer of nearly constant pressure. Its physics is described by the balance between thermal conduction and radiation loss. So a relation can be established between temperature and heat flux which can be scaled according to the pressure. Combining with Shmeleva-Syrovatskii relation as $F_{\rm c}=-\frac{(\kappa_{\rm 0}L(T)T^{3/2})^{1/2}}{\kappa}p$ (Shmeleva \& Syrovatskii 1973), the lower boundary conditions  can be approximated (Meyer et al. 2000a) as,
\begin{equation}
\begin{array}{l}
{T_{\rm i}=T_{\rm e}=10^{6.5}} \,{\rm K},\\
F_{\rm c}=-2.73\times 10^6 P \ {\rm at}\ z=z_0,
\end{array}
\end{equation}

There is no pressure and no heat flux at infinity, which requires
sound transition at some height $z=z_1$. So we constrain the upper
boundary as,
\begin{equation}
F_{\rm c}=0\  {\rm and}
 \  v_{\rm z}=V_{\rm s}\ {\rm at}\ z=z_1.
\end{equation}
With such boundary conditions, assuming an initial value $\lambda$
and a pair of lower boundary values for $P_0$ and $(\rho v_z)_0$, we start
the integration  along $z$. If the trial values for $P_0$ and
$(\rho v_z)_0$ fulfill the upper boundary conditions, the initial value of $P_0$
and $(\rho v_z)_0$ can be taken as the solutions of the differential
equations. The value of $f$ can then be calculated out from Eq.({\ref{equa-f}}).
   The results for a series of $f=0.0,~0.1,~0.2,~0.3,~0.4$ are given in the following section. When $f=0.0$, which means that there are no other energy heating for the corona except the viscous heating, this result is similar to the former works and will be compared with the results for $f > 0.0$.

\subsection{Spectra from the Disk and Corona}
   With the detailed calculations of the corona structure,  the temperature and  density in the corona and the effective temperature are determined.  We then calculate the spectrum from such a disk and corona for given structure.

In the corona, the emissions are from bremsstrahlung radiation and inverse Compton scattering of disk photons.
 For bremsstrahlung cooling spectrum, we adopt the
Eddington approximation and two-stream approximation (Rybicki \&
Lightman 1979; Manmoto et al. 1997) to calculate the radiative flux,
\begin{equation}
F_{\rm \nu}=\frac{2\pi}{\sqrt{3}}B_{\rm \nu}[1-\rm {exp}(-2\sqrt{3}\tau^{*}_{\rm \nu})],
\end{equation}
here $B_{\rm \nu}$ is Planck function, $\tau^{*}_{\rm \nu}=\sqrt{\pi}\kappa_{\rm \nu}H$ is the vertical absorption optical depth. Assuming that the flows is local thermal equilibrium, then $\kappa_{\rm \nu}=\chi_{\rm \nu}/4\pi B_{\rm \nu}$, where$\chi_{\rm \nu}=\chi_{\rm \nu, brems}$ is emissivity of bremsstrahlung (Narayan \& Yi 1995).

Since we focus on HLAGNs,  of which the accretion rate is very high and the disk is the dominant accretion flow,  the soft photons for inverse Compton scatter are supplied by the thin disk
blackbody radiation. However, the ones from bremsstrahlung emission
are neglected because of similar energies to the electrons and cannot gain
much from Compton scattering (Done 2010, Done et al. 2011). We compute the Compton spectrum by
Monte Carlo simulation. The accretion flow consists of a cold disk in the middle and hot
corona above/below. At each radius, the cold disk emits black body
radiation with the temperature $T_{\rm eff}$ as shown in the model
equation (\ref{e:teff}). The soft  photons from the local
disk transverse through the hot corona. Since the hot corona is
optically thin, some of the soft photons pass through it without
scattering, some of them are scattered by the electrons in the corona. We
calculate the energy spectra of photons that emerge from  the corona. The
method of Monte Carlo simulation used in our paper is essentially the same
as that described by Pozdnyakov et al.(1977). In the code, we
introduce the weight $\omega$ to efficiently calculated the effects
of multiple scattering. Firstly, we set initial weight $\omega_{\rm 0}=1$ for a given
soft photon from the cold disk. Then, we calculated the escape
probability $P_{\rm 0}$ of passing through the corona slab. The value of
$\omega_{\rm 0}P_{\rm 0}$ is the transmitted portion and is recorded
to calculate the penetrate spectrum. The remaining weight
$\omega_{\rm 1}=\omega_{\rm 0}(1-P_{\rm 0})$ is the portion that
undergoes more than one scattering. Suppose that $P_{n}$ is the escape
probability after the $n$th scattering, then
$\omega_{\rm n}P_{\rm n}$ is the transmitted portion of photons
after the $n$th scattering. The remaining portion $\omega_{\rm
n}(1-P_{\rm n})$ undergoes the $(n+1)$th scattering. This
calculation is continued until the weight $\omega_{\rm n}$ becomes
sufficiently small. In our calculation, the number of processes
calculated is 100000 and $\omega_{\rm n}< 10^{-4}$. Finally, we
obtain the transmitted spectrum and the soft spectrum.

\section{NUMERIC RESULTS}
To study AGNs, we fix the black hole mass to be  $M_{\rm BH}=10^8 M_{\odot}\rm $ all through our calculations and Eddington accretion rate is $\dot{M}_{\rm Edd}=1.39\times10^{26}~\rm gs^{-1}$. The viscous parameter $\alpha$ in the corona is fixed to be 0.3.

\subsection{The Properties of Corona under Viscous Heating}\label{vicous-properties}
\begin{figure}[htbp]
\begin{minipage}[t]{0.5\linewidth}
\centering
  \includegraphics[scale=0.22]{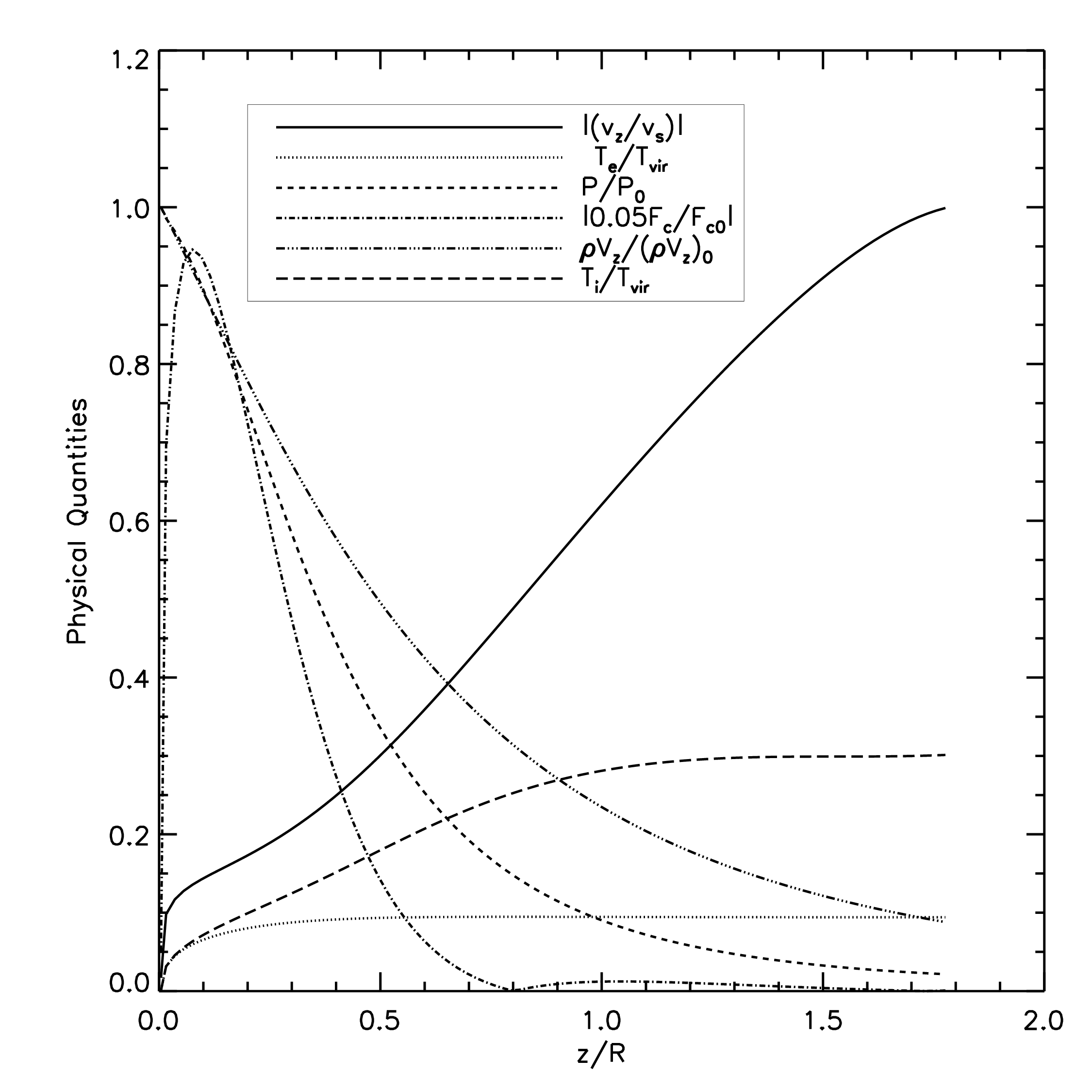}
\end{minipage}
\begin{minipage}[t]{0.5\linewidth}
\centering
\includegraphics[scale=0.22]{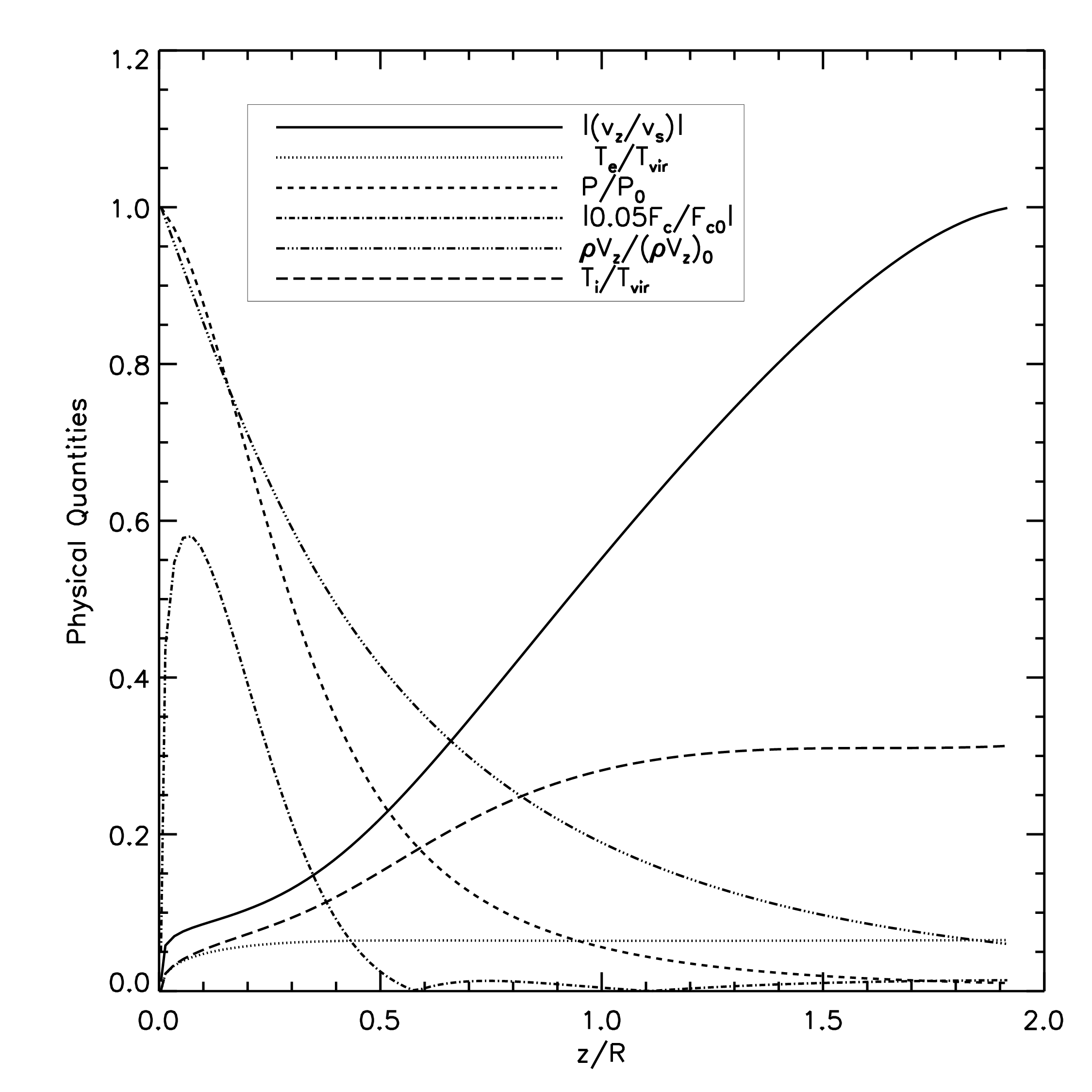}
\end{minipage}
\caption{ Vertical structure of the corona around a black hole at a distance $R=1000~R_{\rm S}$ for accretion rate $\dot{m}=0.02$(left panel) and $\dot{m}=0.2$ (right panel).
Here, $z$ is the height above the equatorial plane of the accretion flow; $T_{\rm e}$ and $T_{\rm i}$ are the
electron temperature and ion temperature, respectively, while $T_{\rm vir}$ is the virial temperature used for scaling. $F_{\rm c}$, $\rho v_{\rm z}$, $P$, and $v_{\rm z}/v_{\rm s}$ are the heat flux, vertical mass flow rate per unit area, total pressure, and vertical velocity scaled by the sound speed, respectively. The quantities at the lower boundary are marked with subscript 0.}
\label{fig:f=0.0 structure}
\end{figure}

 First, we set $\lambda =0.0$, meaning $f=0$, which is the case
that the corona is heated by only viscous friction. We calculate the corona structures
at $R=1000~R_{\rm s}$ for $\dot{m}=0.02$ and $\dot{m}=0.2$ as shown
in the Fig.\ref{fig:f=0.0 structure}, here $\dot{m}=\dot{M}/\dot{M}_{\rm Edd}$ which is the total accretion rate. Comparing the results in two panels,
 we find that the vertical profiles of the coronal quantities are similar
for different accretion rates. From the lower boundary upwards, the pressure ($P$) and the
vertical mass flow ($\rho v_{z}$) decrease dramatically in a thin layer. The downward heat flux ($-F_{c}$) increases
in this layer until a maximum is reached and then decreases along $z$ and can be
neglected on the upper boundary. The temperatures of electrons ($T_{\rm e}$) and ions ($T_{i}$), starting at a coupled value of  $10^{6.5}~\rm K$
at coronal lower layer, increase with the height and decouple at
about $z/R\sim0.1$. Upon this height, $T_{\rm i}$ still increases up
 to $T_{\rm i}\sim 0.3 T_{\rm vir}$ and $T_{\rm e}$ keeps almost the same value
  ($\sim 0.05T_{\rm vir}$) throughout the corona, where $T_{\rm vir}=GMm_{\rm p}\mu/\kappa R=3.37\times10^{12}(R/R_{s})^{-1}~\rm K$ is the virial temperature. These typical features in
  disk corona, which are also shown in earlier works (e.g. Meyer et al. 2000b),
  indicate that the corona above a thin disk undergoes very
  steep changes in temperature and density and cannot be simply averaged in the vertical direction.

\begin{figure}[htbp]
  \vspace{2mm}
\begin{center}
 \hspace{3mm}
 \includegraphics[width=3.5in ,height=3.0in,angle=0.0]{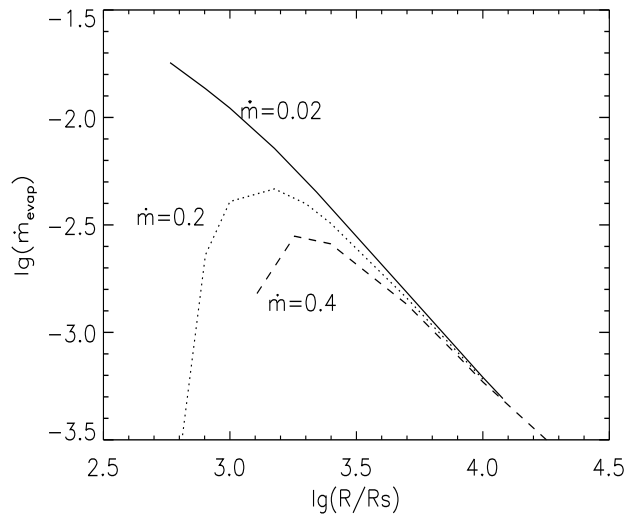}

\caption{The relation between the corona accretion rate $\dot{m}_{\rm evap}$ and the radius for different accretion rate with $f=0.0$. }
   \label{Fig:evap-rate-huadd0.0}
  \end{center}
\end{figure}

We also calculate the evaporation rate $\dot{m}_{\rm evap}$ ($=\dot{M}_{\rm evap}/\dot{M}_{\rm Edd}$) along distances, which represents  the mass flowing rate in the corona. The results are shown in
 Fig.\ref{Fig:evap-rate-huadd0.0}. Similar to the
  previous works (e.g. Liu at al. 2002a), the coronal accretion rate
   increases toward the central black hole, reaches a maximum at a few hundred
Schwarzschild radius and then drops very quickly in the inner region because
of strong inverse Compton scattering, i.e. a large fraction of the corona gas
condensing onto the disk near the black hole. Comparing the
three evaporation curves for accretion rate,  $\dot{m}$=0.02,~0.2 and 0.4, we find that the maximum evaporation
rate decreases with increasing accretion rate and the corona gas begins condensing at outer regions for higher accretion rate. The temperature in electrons, as shown in Fig.1, is lower at higher accretion rates. This indicates that the corona becomes very weak at high accretion rates, like in HLAGNs.

   We can roughly estimate the radiation from the disk and
   corona under the viscous heating. As shown in Fig.\ref{Fig:evap-rate-huadd0.0},
    the maximum corona accretion rate is
   $\dot{m}_{\rm evap,max}=4\times10^{-3}$  at $\sim1250~R_{\rm s}$ for
   $\dot{m}=0.2$. Supposing that the corona keeps accreting at this
   accretion rate, the radiation strength of corona relative to the disk is
  \begin{equation}\label{eq:lcmodld}
  \frac{L_{\rm c}}{L_{\rm d}}= \frac{\eta\dot{M}_{\rm c}c^2}{\eta\dot{M}_{\rm d}c^2}  \la \frac{\dot{m}_{\rm evap,max}}{\dot{m}-\dot{m}_{\rm evap,max}}=0.02.
    \end{equation}
  Here, we have supposed that radiative efficiency $\eta$ in corona is equal to that in disk. From Eq.(\ref{eq:lcmodld}), we can see that the corona is very weak and the radiation is dominated by the disk. In fact, the luminosity ratio between the corona and disk is much smaller than above value since more and more coronal gas condenses into the disk at smaller distance ($R<1250R_{\rm s}$).
   In order to produce strong X-ray in corona, we
  consider that some fraction of gravitational energy is transferred to the corona, providing additional heating mechanism against the strong Compton cooling.

\subsection{The Properties of Corona with Additional Heating}
\subsubsection{Vertical structure}

Assuming a fraction $f$ (=0.1,~0.2,~0.3 and 0.4) of accretion energy is added to the corona, we perform numerical calculations for accretion rates of $\dot{m}=$0.1, 0.2 and 0.5.

\begin{figure}[htbp]
\begin{minipage}[t]{0.5\linewidth}
\centering
\includegraphics[scale=0.22]{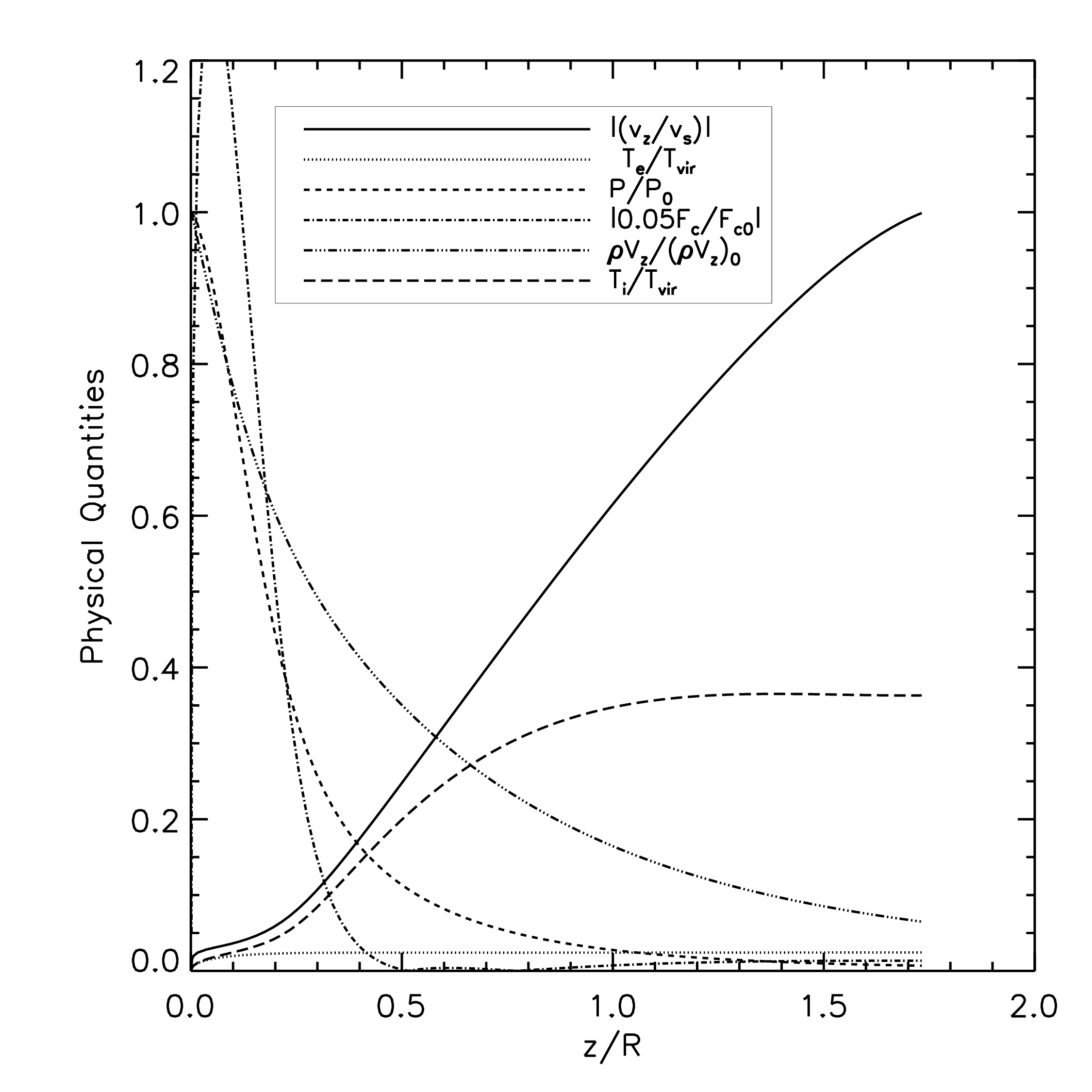}
\end{minipage}
\begin{minipage}[t]{0.5\linewidth}
\centering
\includegraphics[scale=0.22]{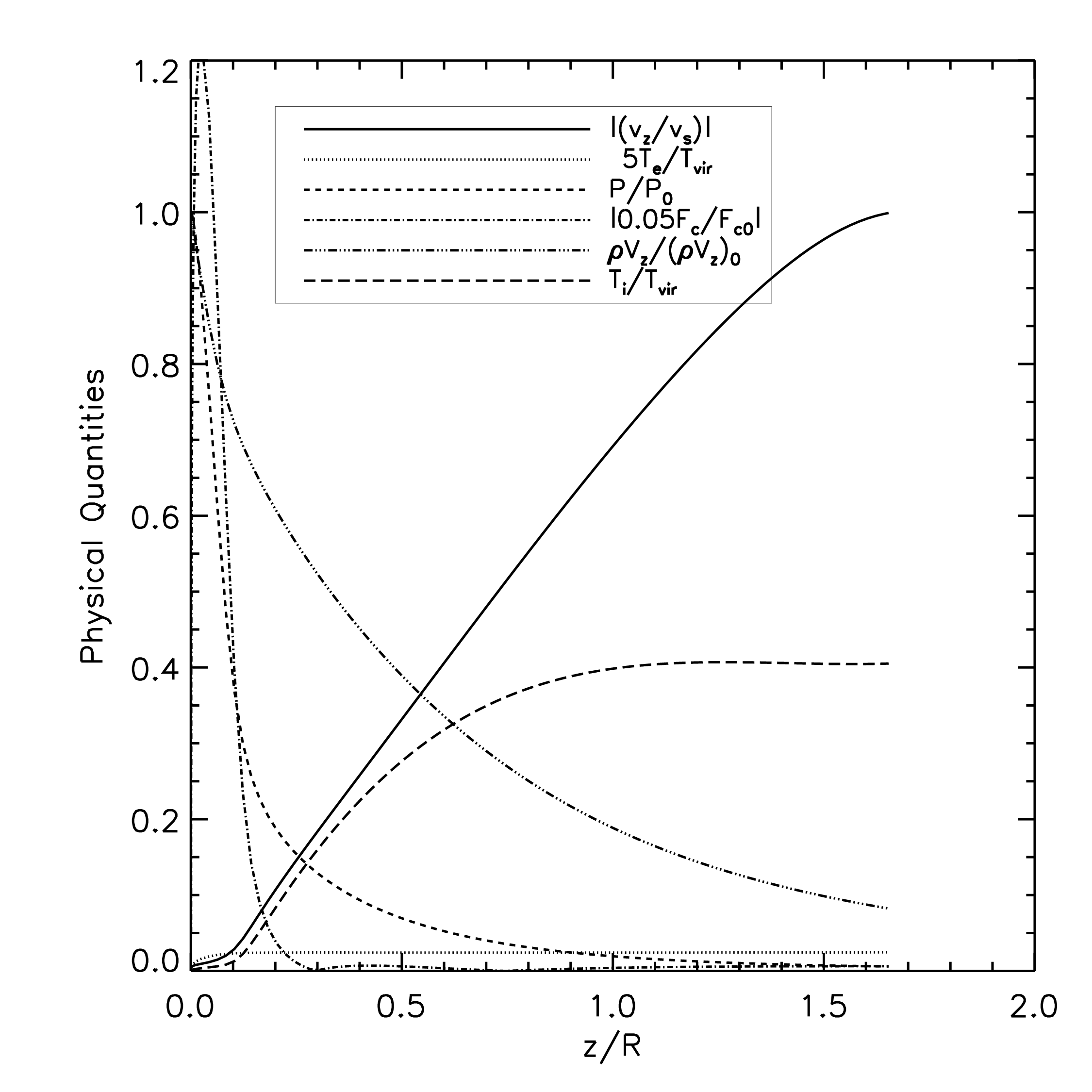}
\end{minipage}
\caption{ The vertical structure of corona for $\dot{m}=0.2,~f=0.2$ at $R=100R_{\rm s}$(left) and $R=20R_{\rm s}$(right).}
\label{fig:f=0.2 structure}
\end{figure}

   We start our calculations from the distance of $100R_{\rm s}$ to the ISCO region ($R= 3.1R_{\rm s}$ for convenience).  It has been shown in section \ref{vicous-properties} that the maximum corona accretion
   rate can only reach $\sim4\times10^{-3}\dot{M}_{\rm Edd}$ at $\sim1000R_{\rm s}$ for $\dot{m}=0.2$. In the inner region, strong Compton cooling leads to condensation of corona. Whether the corona can exist at distances inside $100R_s$ depends on how much accretion energy is added to the corona.  From our test calculations, we find that  for $f=0.2$, a corona can still survive in the region from
  $100R_{\rm S}-3.1R_{\rm S}$.
   The coronal vertical structure is shown in Fig.\ref{fig:f=0.2 structure}.
   The profile is similar to the case for $f=0$ at large distances, though the absolute values are different.


\begin{figure}[htbp]
\begin{minipage}[t]{0.5\linewidth}
\centering
\includegraphics[width=3.5in,height=3.0in]{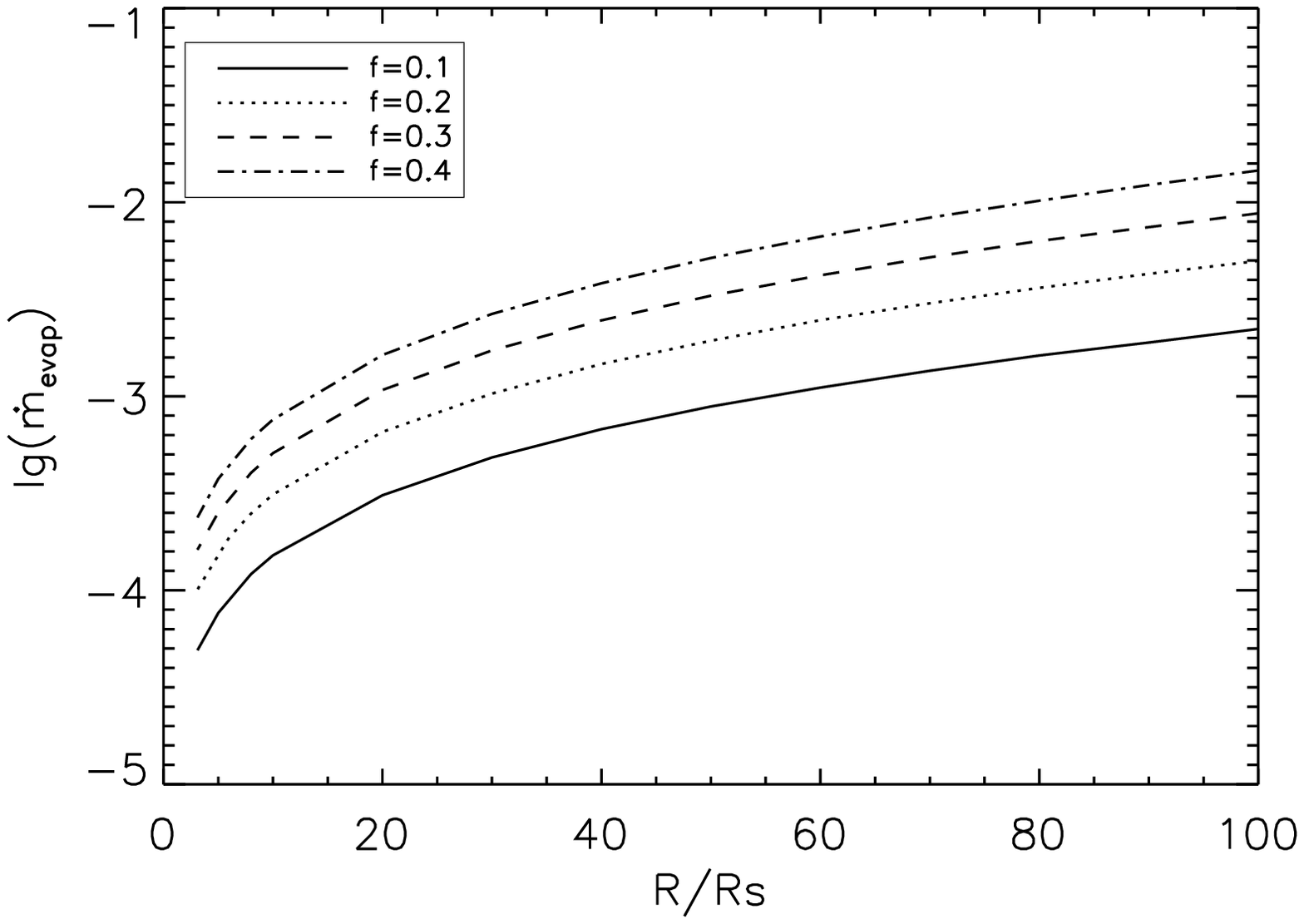}
\end{minipage}
\begin{minipage}[t]{0.5\linewidth}
\centering
\includegraphics[width=3.5in,height=3.0in]{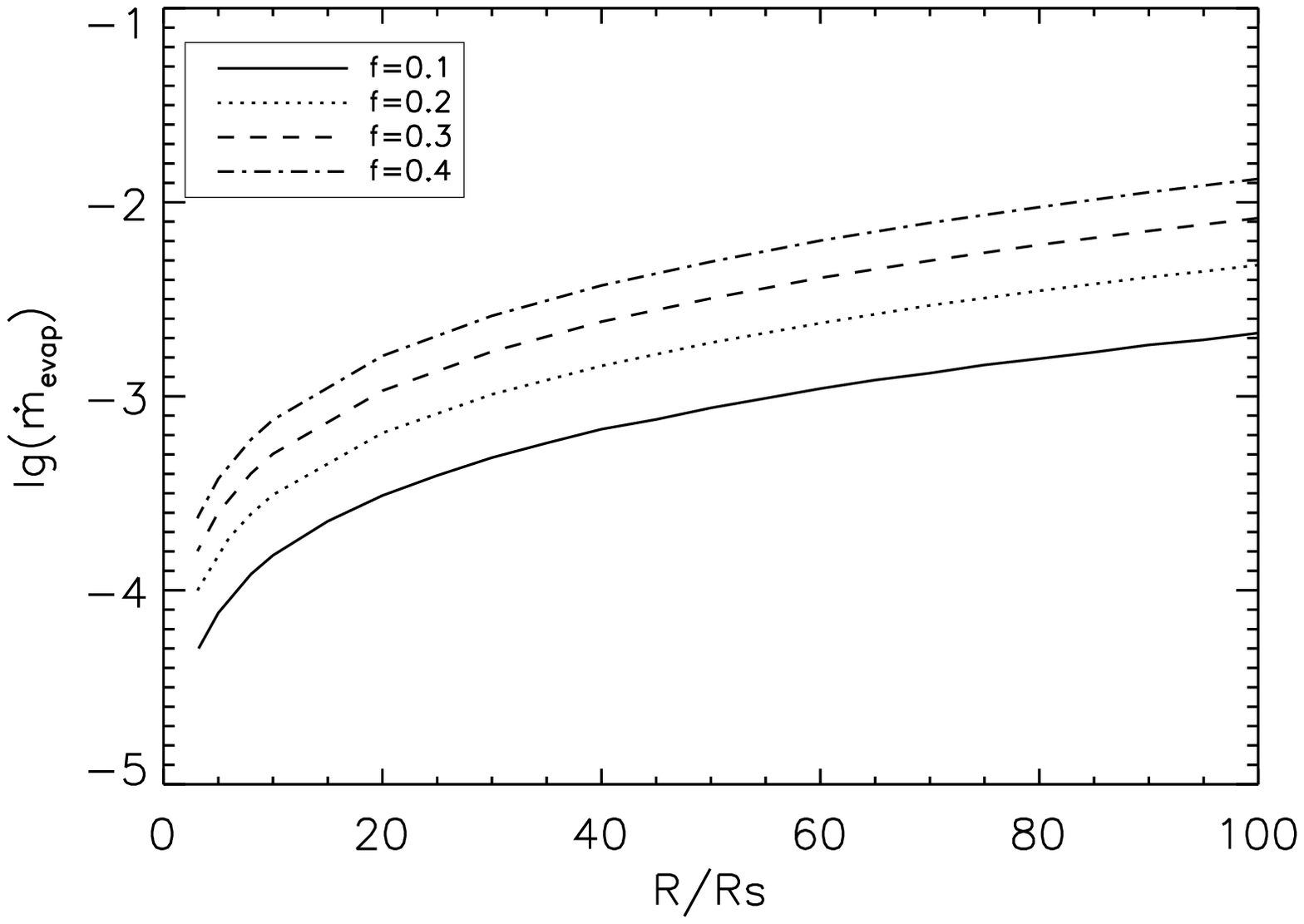}
\end{minipage}
\caption{The radial distribution of corona accretion rate for $\dot{m}=0.2$ (left) and $\dot{m}=0.5$ (right) with different $f$.}
\label{fig:evap-mdot-0.2-0.5}
\end{figure}

\begin{figure}[htbp]
\begin{minipage}[t]{0.5\linewidth}
\centering
\includegraphics[width=3.2in,height=3.0in]{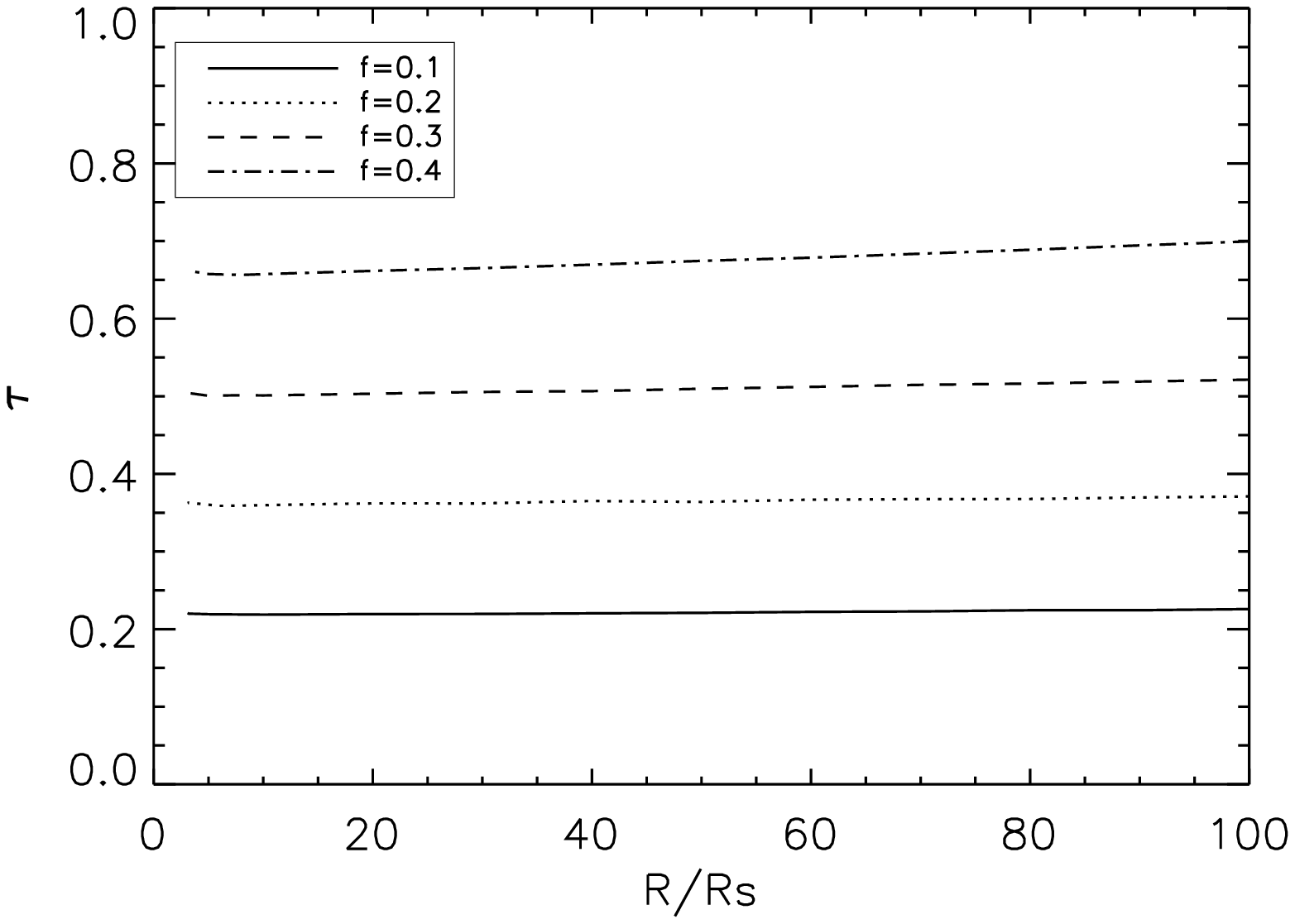}
\end{minipage}
\begin{minipage}[t]{0.5\linewidth}
\centering
\includegraphics[width=3.2in,height=3.0in]{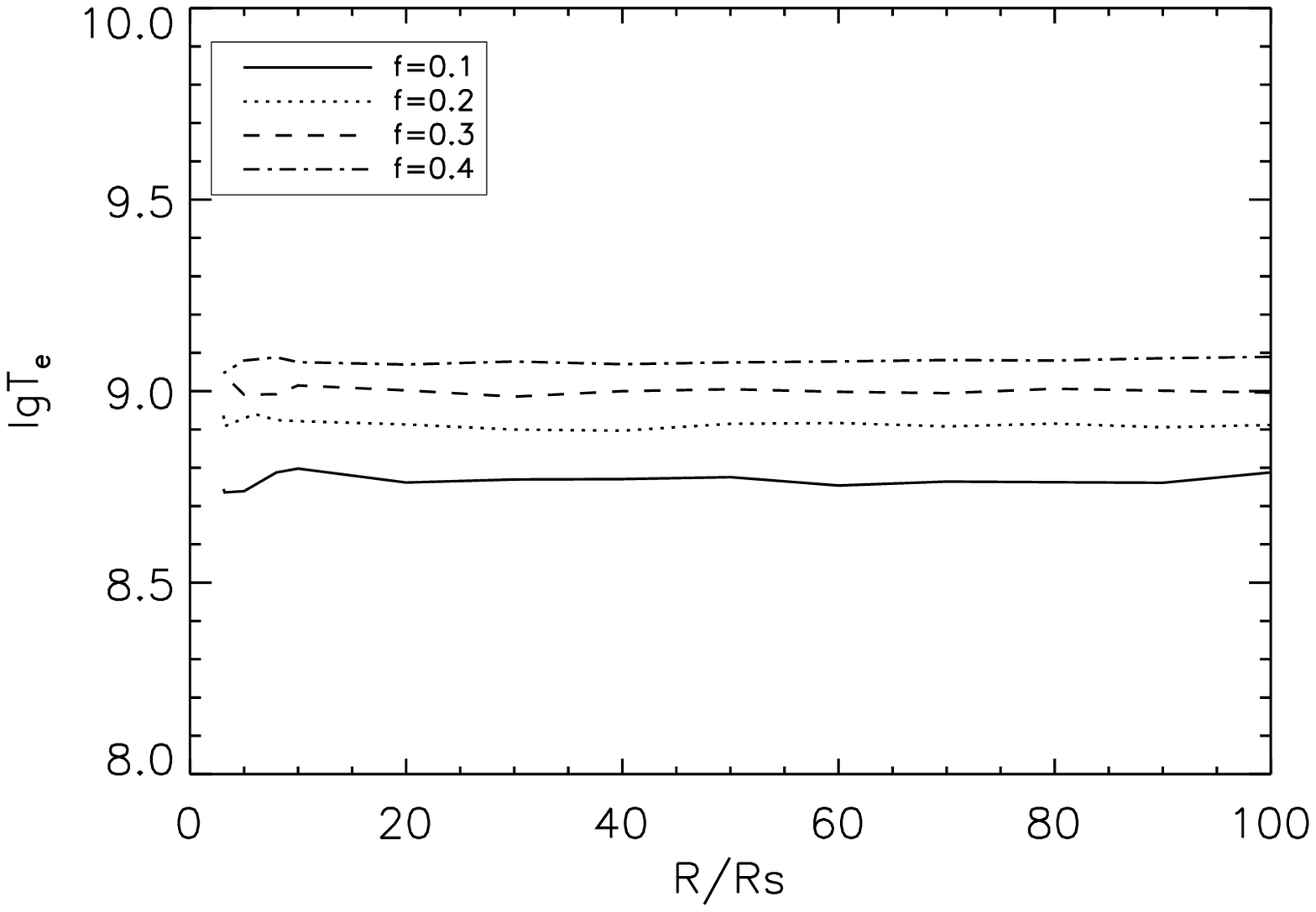}
\end{minipage}
\caption{The radial distribution of scattering optical depth  ($\tau$) and electron temperature ($T_{\rm e}$) for $\dot{m}=0.2$ with different $f$.}
\label{fig:parameter-mdot0.2}
\end{figure}

In order to study the corona properties with different energy input,
we calculate the coronal accretion rate for  $f=0.1,~0.2,~0.3,~0.4$, where
the total mass supply rates are assumed to be  $\dot{m}=0.2,~0.5$
respectively. The results are plotted in Fig.\ref{fig:evap-mdot-0.2-0.5} and Fig.\ref{fig:parameter-mdot0.2}.
One can see from Fig.\ref{fig:evap-mdot-0.2-0.5} that the accretion rate
in the corona decreases with decreasing distance, as a consequence
of condensation. While the electron temperature and scattering optical depth
are nearly the same at distances up to 100$R_{\rm s}$, as shown in
Fig. \ref{fig:parameter-mdot0.2}. Besides, Fig. \ref{fig:evap-mdot-0.2-0.5}
also shows that  the larger fraction $f$ is, the larger becomes the gas flow in the corona.
This is because more energy goes directly to heat the electrons
in corona to a higher temperature, leading to more conductive flux
and hence more evaporation or less condensation. Thus, more gas is
kept in and accreted through the corona. This is confirmed by the
higher electron temperature and larger scattering optical depth for larger $f$,
as shown in Fig. \ref{fig:parameter-mdot0.2}.

  Comparing with the two panels in Fig.\ref{fig:evap-mdot-0.2-0.5}, we understand that the coronal accretion rates are insensitive to the mass supply rate, as long as the fraction of accretion energy released in the corona is the same. Calculations show that the electron temperature and scattering optical depth of the corona do not vary with the accretion rate either.  This can be understood as a consequence of energy balance in the corona. For given $f$, when the mass supply rate  ($\dot m$) increases, the heating rate added to the corona also increases ($\propto f\dot m$). On the other hand, the cooling rate by inverse Compton scattering also increases with the increasing density of soft photon ($\propto (1-f)\dot m$, if $(1-f)\dot m>>\dot m_{\rm evap}$). If the additional heat and Compton cooling are the dominant energy terms for the coronal electrons, there should be no variation in both the evaporation rate and  Compton $y$-parameter. Therefore, the electron temperature does not vary either.

\subsubsection{Spectra of the disk and corona}

Deriving the structure of our model at each radius, we get
radius-dependent electron temperature and electron scattering optical depth,
which are the input parameters for bremsstrahlung and Compton spectrum. The bremsstrahlung radiation is proportional to $n^2_{\rm e}$ but the inverse Compton scattering is proportional to $n_{\rm e}$. So from the  vertical structure, we can see that the bremsstrahlung dominates in the lower layer of the corona and the upper layer is dominated by the Compton radiation.
Since the temperature and number density in the upper corona still depend
on the vertical height, we need to choose appropriate temperature and density at every given distance for calculating the Compton spectrum with Monte Carlo simulation.   Here we constrain them by  the luminosity, i.e., when the luminosity derived from the Monte Carlo simulation equals to the one derived from the coronal structure calculation, the chosen temperature and density are thought to be right and we take this spectrum as the true  spectrum from the corona.

  \begin{figure}[htbp]
\begin{minipage}[t]{0.5\linewidth}
\centering
\includegraphics[width=3.2in,height=3.0in]{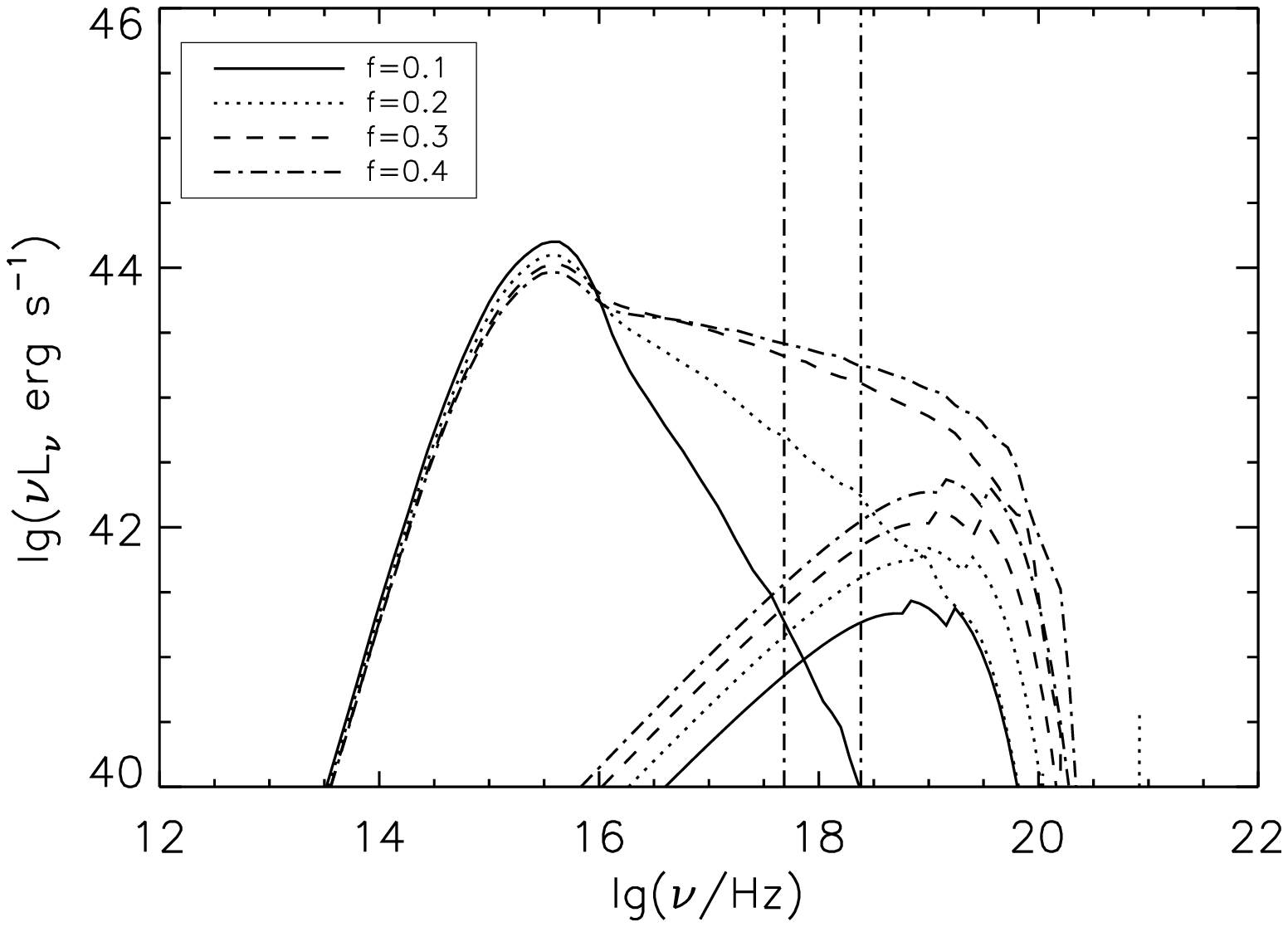}
\end{minipage}
\begin{minipage}[t]{0.5\linewidth}
\centering
 \includegraphics[width=3.2in,height=3.0in]{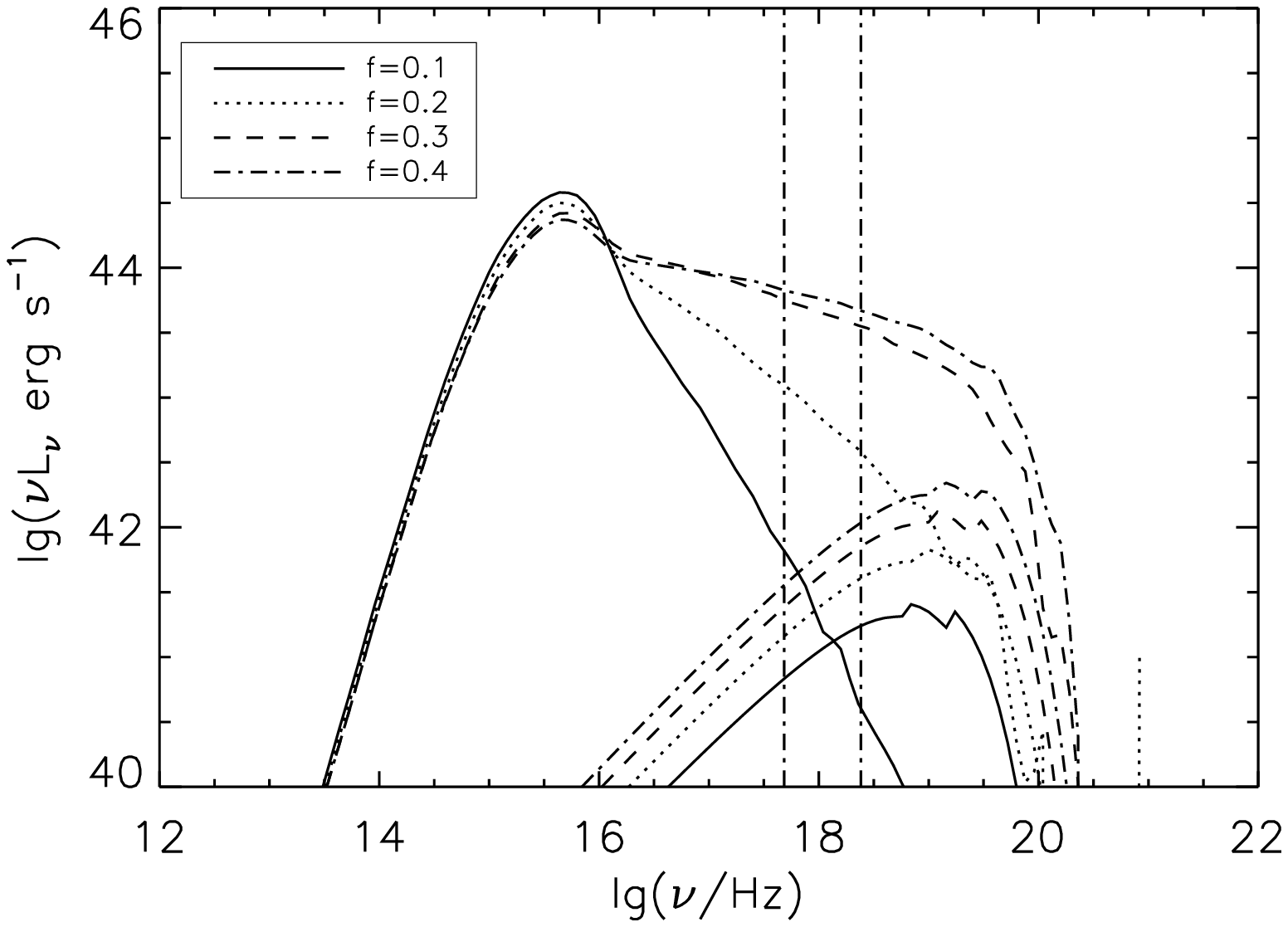}
\end{minipage}
\caption{The bremsstrahlung and Compton spectrum for $\dot{m}=0.2$ (left) and $\dot{m}=0.5$ (right). The curves with the bump in about $10^{19}$ Hz respond to the bremsstrahlung emission, and the ones with bump in the UV range are contributed by the blackboday radiation in the disk and inverse Compton scattering in the corona. The two vertical dotted lines indicate the photon energies of 2 keV and 10 keV, respectively. }
\label{fig:nu-lnu-mdot0.2-0.5}
\end{figure}

\begin{figure}[htbp]
\begin{minipage}[t]{0.5\linewidth}
\centering
\includegraphics[width=3.2in,height=3.0in]{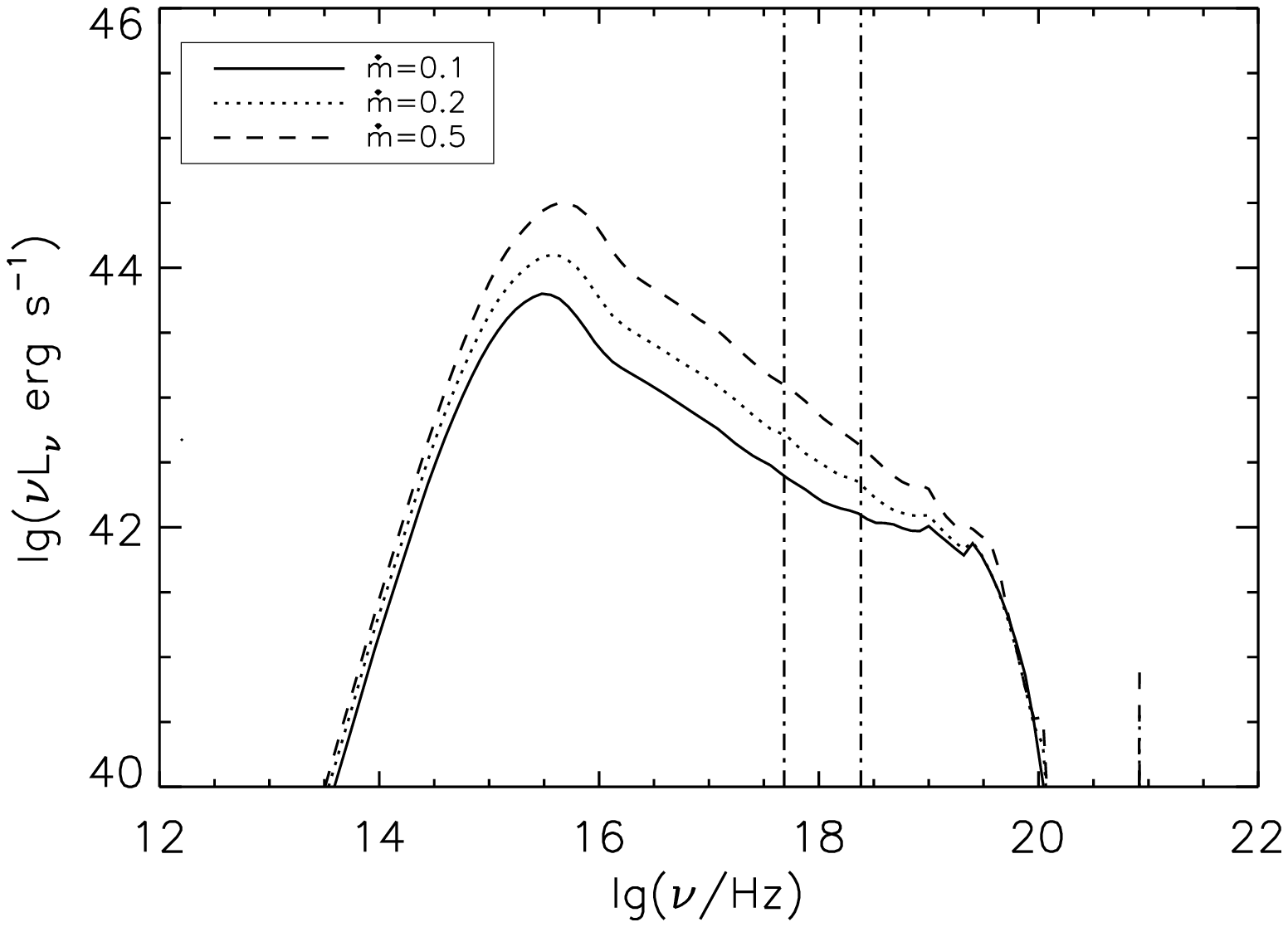}
\end{minipage}
\begin{minipage}[t]{0.5\linewidth}
\centering
\includegraphics[width=3.2in,height=3.0in]{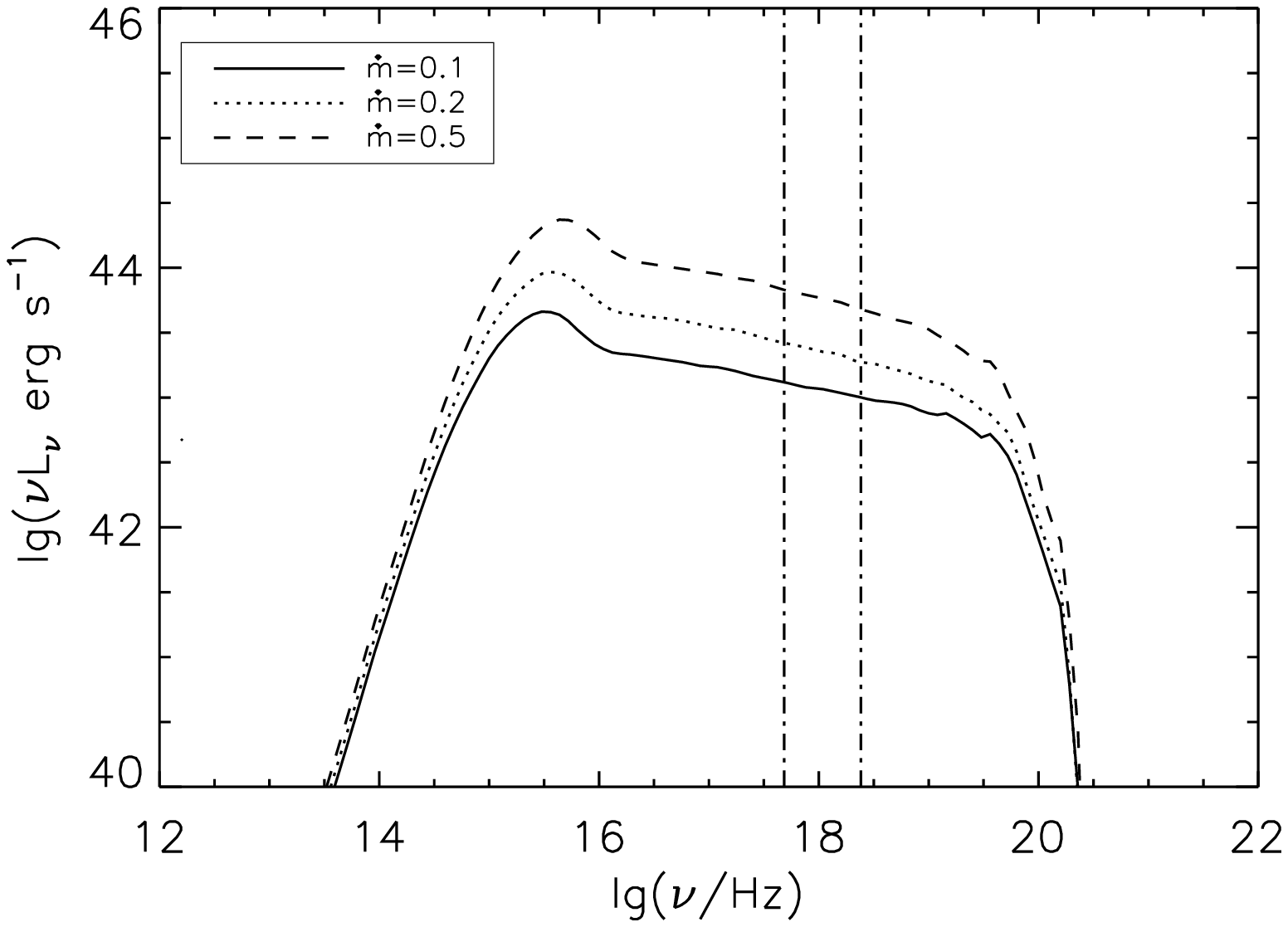}
\end{minipage}
\caption{The total spectrum contributed from multi-blackbody from disk, Bremsstrahlung and inverse Compton scattering for $f=0.2$ (left panel), $f=0.4$ (right panel).}
\label{fig:nu-lnu-f0.2-0.4}
\end{figure}

  Fig.\ref{fig:nu-lnu-mdot0.2-0.5} shows the spectra for different $f$, $f=0.1,~0.2,~0.3,~0.4$, and different accretion rates, $\dot{m}=0.2, ~0.5$. The spectrum is Compton dominated in the high accretion rate. The larger $f$ is, the stronger becomes both Compton and bremsstrahlung radiation, since the coronal density and temperature increase with increasing heating to the corona; While the optical-UV radiation, which is from the disk, slightly decreases with increasing $f$. The spectrum of hard X-ray in the range of $2-10~\rm keV $ is marked by the two vertical dashed lines. The hard X-ray spectrum is harder for higher $f$ and $L_{\rm bol}/L_{\rm 2-10~\rm keV}$ decreases with $f$. For $f=0.1$, the spectrum seems so steep that the corona radiation is still very weak. The photon index for the hard X-ray is in the range of $2.2-2.7$ for $0.2\le f\le 0.4$, which roughly fits the spectrum of HLAGNs.  Table \ref{table:spectrum index} lists photon index for $2-10~\rm keV$ and the ratio of $L_{\rm bol}/L_{\rm 2-10~\rm keV}$. For different accretion rates, the spectrum index is nearly the same with the same $f$, which can also be seen from the total spectrum shown in Fig. \ref{fig:nu-lnu-f0.2-0.4}. Comparing with the two panels in the Fig.\ref{fig:nu-lnu-f0.2-0.4}, we understand that the spectrum varies with the fraction of accretion energy liberating in the corona, while it does not sensitively depend on $\dot{m}$, as long as $f$ is independent on the accretion rate. This can be understood from our structure calculating results, namely,  for given $f$ the electron temperature in corona $T_{\rm e}$, and the scattering optical depth $\tau$, do not change as the accretion rate, though the disk radiation increases and hence Compton  radiation also increases with increasing accretion rate.

\begin{figure}[htbp]
\begin{center}
\includegraphics[height=3.2in,width=3.5in]{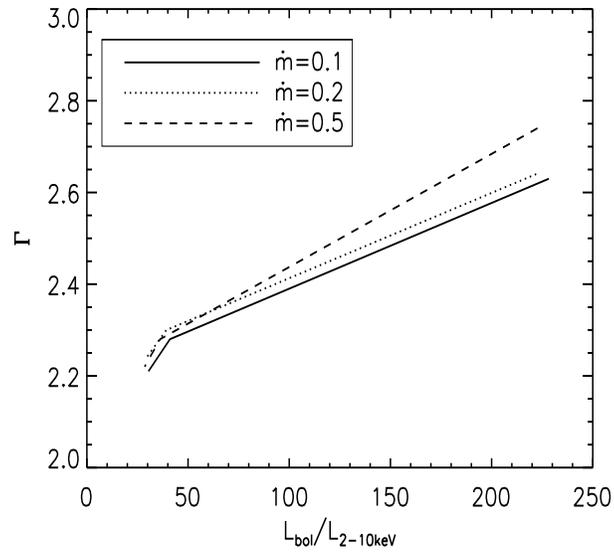}
\caption{ The relation between photon index ($\Gamma$
 ) and bolometric correction ($L_{\rm bol}/L_{\rm 2-10~keV}$). It shows that $\Gamma$ increases with $L_{\rm bol}/L_{\rm 2-10~keV}$, which is a consequence of decreasing energy fraction released in the corona.}
\label{fig:gamma-bol-xray}
 \end{center}
\end{figure}

The photon index and the luminosity ratio, as listed in Table 1,  indicate that a  fraction of accretion energy  is necessary in order to explain observations in HLAGNs.  For $0.2\le f\le 0.4$, the predicted photon index is around $2.2-2.7$, and the bolometric correction from hard X-rays $L_{\rm bol}/L_{\rm 2-10~\rm keV}$ is in the range of  $28-228$.  These are roughly consistent with observations (Vasudevan \& Fabian 2007, 2009; Shemmer et al. 2006 and references therein; Zhou \& Zhao 2010).  In Fig. \ref{fig:gamma-bol-xray} we show how the photon index in X-rays varies with bolometric luminosity.  The figure reveals that the  bolometric correction is larger for a steeper X-ray spectrum, which seems to be a common feature for objects with different Eddington ratios.  This is a consequence of  decrease of energy fraction released in the corona.
We now raise the questions as, what is the underlying physics driving the accretion energy to release in the corona? Does the energy fraction $f$ depend on the accretion rate or the black hole mass? We discuss this in the following section.


\begin{table}[htbp]
\begin{center}
\caption{The photon index and $L_{\rm bol}/L_{\rm 2-10~\rm keV}$ for different accretion rate with different $f$.} \label{table:spectrum index}
\begin{tabular}{|c|c|c|c|c|c|c|}
\multicolumn{7}{c}{}\\
  \hline
 & \multicolumn{2}{|c|}{$\dot{m}=0.1$} & \multicolumn{2}{|c|}{$\dot{m}=0.2$}  & \multicolumn{2}{|c|}{$\dot{m}=0.5$} \\
     \hline
    $f$ & $\Gamma$ & $L_{\rm bol}/L_{\rm 2-10~\rm keV}$ & $\Gamma$ & $L_{\rm bol}/L_{\rm 2-10~\rm keV}$ & $\Gamma$ & $L_{\rm bol}/L_{\rm 2-10~\rm keV}$ \\
    \hline
0.2 & 2.6   & 228  & 2.6   & 222  & 2.7   & 222   \\
0.3 & 2.3 & 41  & 2.3   & 39  & 2.3  & 36  \\
0.4 & 2.2  & 30  & 2.2   & 29  & 2.2  & 28  \\
  \hline
\end{tabular}
\end{center}
\end{table}

\section{DISCUSSION}
\subsection{Heating Mechanism in the Corona---Magnetic Heating?}
It has been a long history in studying  magnetic field in the accretion flow,  either  with magnetohydrodynamics(MHD) simulations (e.g. Balbus \& Hawley  1998; Matsumoto 1999; Stone \& Pringle 2001; Machida et
al. 2001; Machida \& Matsumoto 2003; Kato et al. 2004; Kawanaka et al. 2008 and references therein; Ohsuga, et al. 2009, 2011; Penna et al. 2010) or numerical simulation to interpret observations (Liu, et al. 2002b, 2003; Cao 2009).

  There are good reasons to believe that  the magnetic field plays an important role in transporting the accretion energy from the disk to the corona (Haardt \& Marashi 1991).
  Liu et al. (2002b; 2003) investigated the Parker instability in the accretion flow, following the model for the solar corona (e.g. Yokoyama \& Shibata 2001) and found that the accretion energy can be transported to the corona and heats the electrons in the corona. In the frame of disk corona with mass evaporation,  Qian, Liu \& Wu (2007) analyzed the coronal structure and evaporation features with the magnetic field and found that the maximum evaporation rate stays more or less constant ($\sim 3\%~\dot{M}_{\rm Edd} $ ) when the strength of the magnetic field changes, but that the radius corresponding to the maximum evaporation rate decreases with increasing magnetic field. Qiao \& Liu (2009) investigated the influence of viscosity parameter $\alpha$ in the corona and suggested that the coronal radiation is largely enhanced when viscosity increases.

 However, it is not very clear how magnetic field connects with viscosity in the evaporation-fed corona, though some investigations of accretion flows show that the magnetic field and viscosity are associated (e.g. Balbus \& Hawley 1991; Matsumoto \& Tajima 1995; Abramowicz et al. 1996; Balbus 2003).
Here we discuss whether the predicted spectrum can be consistent with observations if it is the magnetic field that brings a fraction of gravitational energy to the corona.

  Let us assume that the magnetic field is generated in the disk by dynamo action. Then the magnetic loops can emerge out of the disk by Parker buoyant instability. In this process, the accretion energy stored in the magnetic field is transported into the corona and thereby liberated through magnetic reconnection. According to the equipartition theorem between the magnetic pressure and gas pressure, we set $P_{\rm B}=\frac{B^2}{8\pi}=\frac{1}{\beta} P_{\rm g,d}$, where $P_{\rm B}$ and $P_{\rm g,d}$ are magnetic pressure and gas pressure in the disk. Then the magnetic energy flux is
\begin{equation}
F_{\rm B}=\frac{B^2}{4\pi}\sqrt{\frac{B^2}{4\pi\rho_{\rm d}}}=(2P_{\rm B})^{\frac{3}{2}}\rho_{\rm d}^{-\frac{1}{2}}=(\frac{2}{\beta}P_{\rm g,d})^{\frac{3}{2}}
    \rho_{\rm d}^{-\frac{1}{2}}.
 \end{equation}

If this part of magnetic energy flux  is the additional heating  source for the corona, the ratio of this part of energy to total gravitational energy is
\begin{equation}\label{fraction}
f=\frac{F_{\rm B}}{F_{\rm tot}}=(\frac{2}{\beta}P_{\rm g,d})^{\frac{3}{2}}\rho_{\rm d}^{-\frac{1}{2}}/F_{\rm tot},
  \end{equation}
where $F_{\rm tot}$ is the total accretion flux
through accretion process, $F_{\rm tot}=\frac{3GM\dot M}{8\pi R^3 }\left[1-(3R_{\rm
s}/R)^{1/2}\right]=8.56\times10^{26}\dot{m}m^{-1}r^{-3}\Phi$, where $\Phi=1-\left(\frac{3}{r}\right)^{\frac{1}{2}}$.

 In the thin disk, the pressure of the disk is
 \begin{equation}\label{e:state-disk}
  P=P_{\rm g,d}+P_{\rm B}+P_{\rm r,d}=(1+\frac{1}{\beta})\frac{\rho_{\rm d} k T_{\rm d}}{\mu m_{\rm p}}+\frac{aT_{\rm d}^4}{3},
 \end{equation}
 where $\mu=0.62$ is the averaged molecular weight. Since some fraction ($f$) of gravitational energy is carried into the corona by magnetic activity, then the energy equation in the disk is
 \begin{equation}\label{e:energy-disk}
 \frac{4ac T_{\rm d}^4}{3\tau}=(1-f)\times F_{\rm tot}=(1-f)\frac{3GM\dot{M}}{8\pi R^3 }\left[1-(3R_{\rm s}/R)^{1/2}\right],
\end{equation}
where $\tau=(\kappa_{\rm es}+\kappa_{\rm 0}\rho_{\rm d}T_{\rm d}^{-3.5})\Sigma$, $\kappa_{\rm es}=0.4~\rm cm^2~g^{-1}$, $\kappa_{\rm 0}=6.4\times 10^{22}~\rm cm^2~g^{-1}$ and $\Sigma=2\rho_{\rm d} H_{\rm d}$ is the surface density of the disk. And the angular momentum equation in the disk is
\begin{equation}\label{e:momentum-disk}
\frac{\dot{M}}{3\pi}\Phi=\nu\Sigma,
\end{equation}
where $\nu=\frac{2}{3}\alpha c_{\rm s}H_{\rm d}$, the viscous velocity $c_{\rm s}=\Omega H_{\rm d}$.

If the gas pressure in the disk is dominant over radiation pressure, $P_{\rm g,d}\gg P_{\rm r,d}$,  and Thomson scattering is much higher than free-free absorption, $\sigma_{\rm T}\gg \sigma_{\rm ff}$, we have $\tau \approx \kappa_{\rm es}\Sigma$, which is similar to the case b) in  Shakura \& Sunyeav (1973) except that $r=R/R_{\rm s}$ is taken here instead of  $r=R/3R_{\rm s}$. We can derive the disk quantities. The density in the disk is
\begin{equation}
\rho_{\rm d}=21.43(\alpha m)^{-\frac{7}{10}}\left[\dot{m}\Phi\right]^{\frac{2}{5}}r^{-\frac{33}{20}}(1+\frac{1}{\beta})^{-\frac{6}{5}}(1-f)^{-\frac{3}{10}},
 \end{equation}
and the gas pressure is
 \begin{equation}
P_{\rm g,d}=1.91\times 10^{18}(\alpha m)^{-\frac{9}{10}}\left[\dot{m}\Phi\right]^{\frac{4}{5}}r^{-\frac{51}{20}}(1+\frac{1}{\beta})^{-\frac{7}{5}}(1-f)^{-\frac{1}{10}}.
\end{equation}

 With this disk pressure and density, the energy fraction can be obtained from Eq.(\ref{fraction}),

 \begin{equation}\label{fraction-g}
   f=1.88\alpha^{-1}(1+\beta)^{-\frac{3}{2}}.
   \end{equation}
 One can see that $f$ is independent on the accretion rate if the disk is gas pressure dominated.  This implies that a constant fraction of disk accretion energy is transported to the corona at different accretion rates, if equipartition coefficient $\beta$ does not vary with accretion rates.  If this is the case for the additional heating to the corona,  the model spectrum will not vary with the accretion rate, as shown in Fig.\ref{fig:nu-lnu-f0.2-0.4}.

 In fact, the accretion disk around a supermassive black hole is often dominated by radiation pressure. In the case of $P_{\rm g,d}\ll P_{\rm r,d}$   and $\sigma_{\rm T}\gg \sigma_{\rm ff}$, the density and pressure in the disk are expressed as,
 \begin{equation}
 \rho_{\rm d}=1.14\times10^{-6}(\alpha m)^{-1}\left[\dot{m}\Phi\right]^{-2}r^{\frac{3}{2}}(1-f)^{-3} ,
  \end{equation}
  \begin{equation}
 P_{\rm g,d}=6.21\times10^{9}(\alpha m)^{-\frac{5}{4}}\left[\dot{m}\Phi\right]^{-2}r^{\frac{9}{8}} (1-f)^{-\frac{13}{4}},
   \end{equation}
which are similar to the solution of case a) in  Shakura \& Sunyeav (1973).  Then we can determine $f$ from Eq. (\ref{fraction}) as
 \begin{equation}\label{fraction-r}
   f(1-f)^{\frac{27}{8}}=1.52\times10^{-9}\alpha^{-\frac{11}{8}}m^{-\frac{3}{8}}(\dot{m}\Phi)^{-3}r^{\frac{63}{16}}{\beta}^{-\frac{3}{2}}.
   \end{equation}

\begin{figure}
\includegraphics[width=3.2in,height=3.0in]{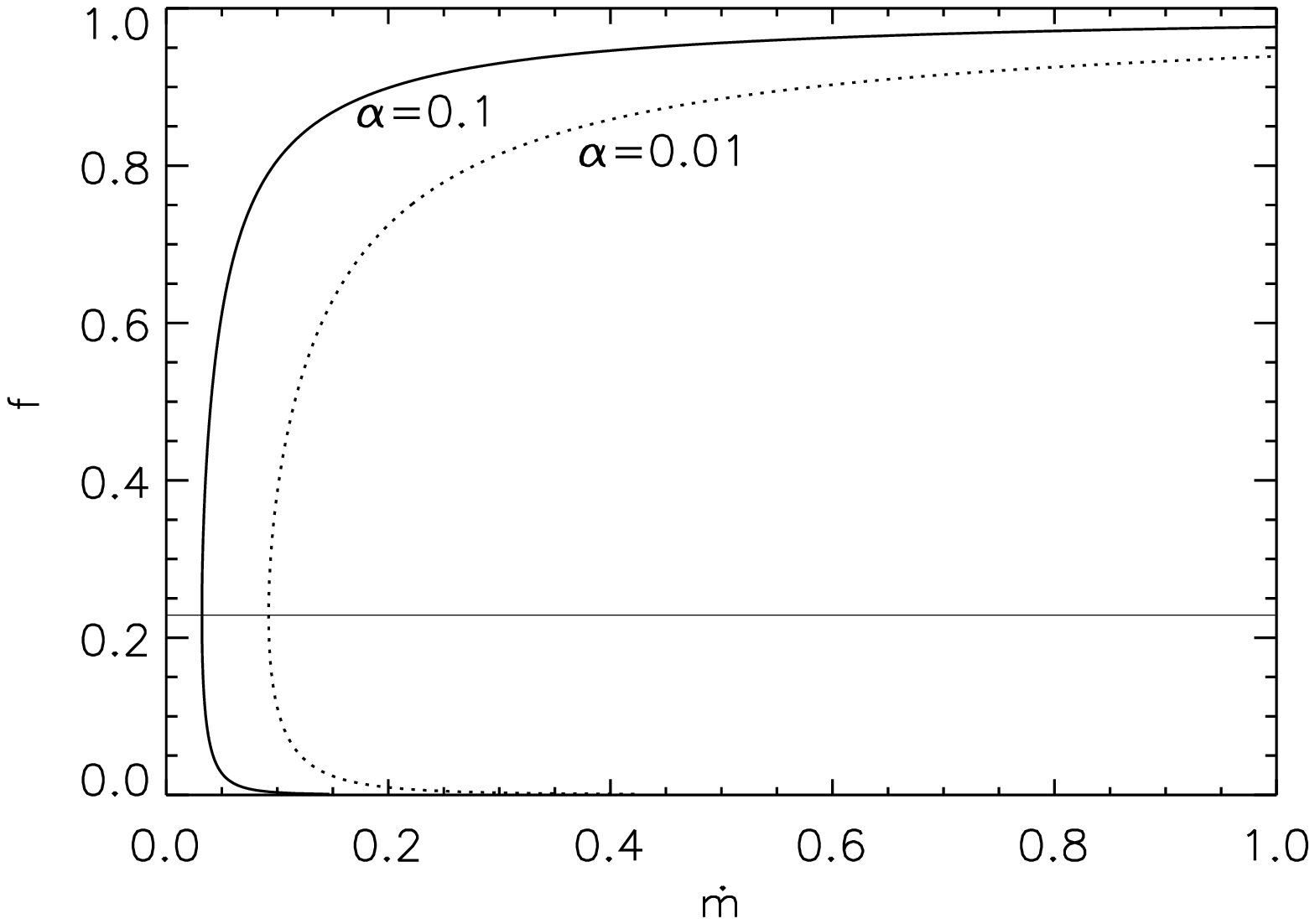}
\includegraphics[width=3.2in,height=3.0in]{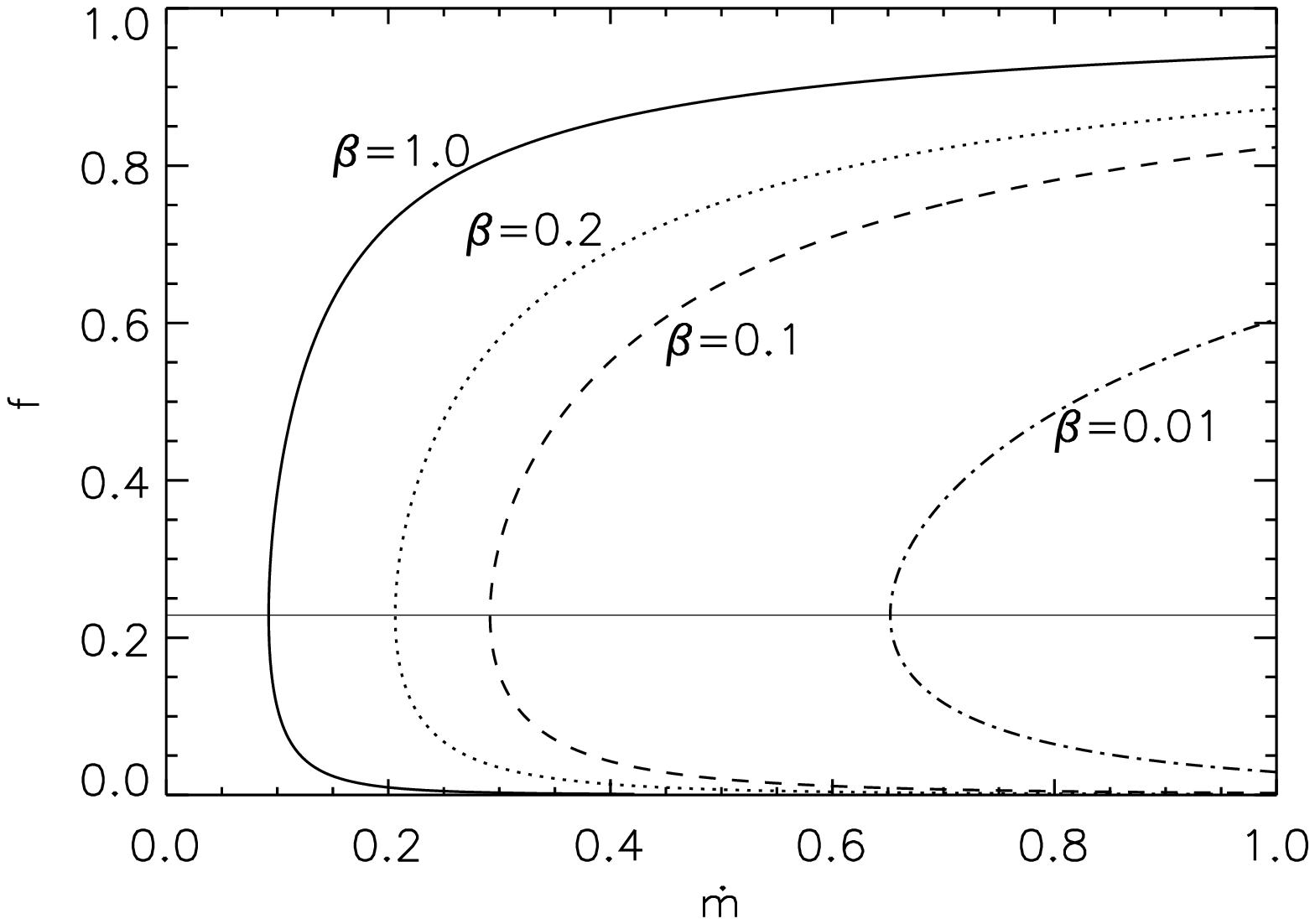}
\caption{The fraction of accretion energy released in the corona as a function of accretion rate. For a given accretion rate, there are two values for $f$, one is $f>\frac{8}{35}$, the other is $f<\frac{8}{35}$. The horizontal line indicates  $f=8/35$, above which $f$ increases with increasing accretion rate and below which $f$ decreases with $\dot m$. The effect of  viscosity is shown in the left panel by assuming $\beta=1$ and the effect of magnetic field is shown in the right panel by assuming $\alpha=0.01$.  It can be seen that for small viscosity and/or strong magnetic field,  the accretion rate corresponding to a certain $f$ is large.}
\label{fig:f-mdot}
\end{figure}

 Apparently, in a radiation pressure dominated disk,  $f$ depends on the accretion rate. The function $f(1-f)^{\frac{27}{8}}$ in Eq. (\ref{fraction-r})  reaches a maximum of $0.095$ at $f=\frac{8}{35}$, which requires $1.52\times10^{-9}\alpha^{-\frac{11}{8}}m^{-\frac{3}{8}}(\dot{m}\Phi)^{-3}r^{\frac{63}{16}}{\beta}^{-\frac{3}{2}} \le 0.095$. This indicates that the accretion rate must be larger than a certain value in order to get a solution for $f$ from Eq. (\ref{fraction-r}). If this condition  $1.52\times10^{-9}\alpha^{-\frac{11}{8}}m^{-\frac{3}{8}}(\dot{m}\Phi)^{-3}r^{\frac{63}{16}}{\beta}^{-\frac{3}{2}} \le 0.095$ is fulfilled,  there are two sets of solutions for $f$ ( We saw this bimodal nature in Liu et al. (2002b)). As shown in  Fig.\ref{fig:f-mdot},  one set of solutions  is in the range of $0<f<\frac{8}{35}$,  where $f$ decreases with increasing accretion rate $\dot m$;  While the other one is in the range of $\frac{8}{35}<f<1$,  where $f$ increases with increasing $\dot m$.
 The  relation between the energy fraction and the accretion rate, i.e. $f(\dot m)$,  depends on the viscosity and magnetic field.  In the left panel of Fig.\ref{fig:f-mdot}, we show the effect of viscosity by assuming $\beta=1$ under the theory of equipartition of gas pressure and magnetic pressure. Actually, the local and globe MHD simulation in radiation-dominated accretion flow reveals that magnetic energy density may exceed gas energy density but kept below radiation energy density, which means that $\beta$ can be less than  1.0  (e.g. Turner et al. 2003; Ohsuga et al. 2009, 2011). The effect of magnetic fields is plotted in the right panel of Fig.\ref{fig:f-mdot}.
 It shows that the influence of magnetic field is significant. The stronger the magnetic field, the larger $\dot{m}$ corresponding to $f=8/35$. For a given accretion rate, for example,  $\dot{m}=0.4$, $f$ is larger with stronger magnetic field (i.e. smaller $\beta$) in the lower part of the curve ($f<8/35$).

\begin{figure}[htbp]
\centering
 \includegraphics[width=4.0in,height=3.5in]{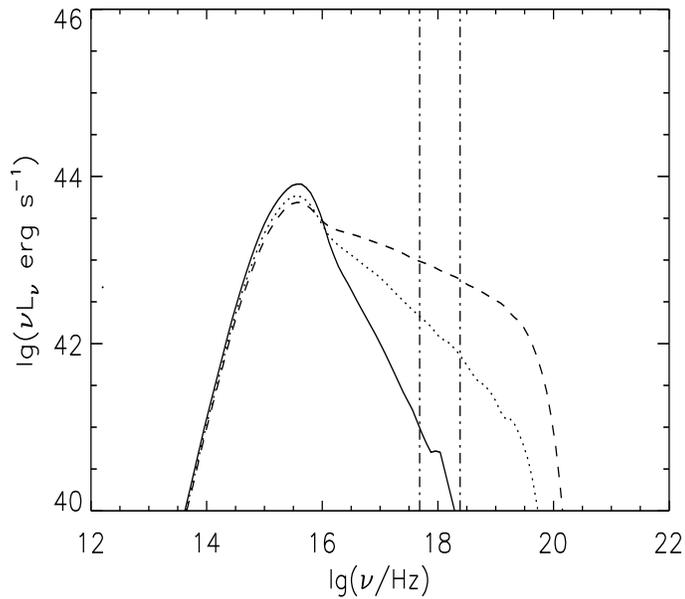}
\caption{The spectrum for different accretion rate. The dashed lines represent a spectrum for a low accretion rate, the dotted line indicates a higher accretion rate, and the solid line is for much higher accretion rate. The fraction of energy released in the corona for theses three curves corresponds to $0.3$, 0.2 and 0.1, respectively. The corresponding $\dot{m}$ is determined  by Eq.(\ref{fraction-r}) by assuming disk viscous parameter $\alpha=0.01$, $\beta=1$ for $m=10^8$ black hole at the most efficient radiation region $r=49/12$.}
\label{fig:spectra-mdot}
\end{figure}

Observed spectral features provide a clue to possible range of $f$.  If  $f<\frac{8}{35}$,  the fraction of accretion energy transferred to the corona decreases with accretion rate. As a consequence, the hard X-ray spectrum becomes softer at high accretion rates,  as shown  in Fig.\ref{fig:spectra-mdot}.
However,  if a large fraction of accretion energy ($f>\frac{8}{35}$) is transferred to the corona, $f$ increases with accretion rate, consequently, the spectrum is harder at higher accretion rates.  Comparing the model prediction and observations,  a small fraction of accretion energy liberation in the corona is preferred, that is, $<\frac{8}{35}$. In Fig.\ref{fig:spectra-mdot}, we plot the spectrum for different accretion rate. The dashed lines represent a spectrum for a low accretion rate, and the dotted line is for a relatively higher accretion rate, and the solid line is for a much higher accretion rate, where the fraction of energy released in the corona corresponds to 0.3, 0.2, 0.1 respectively, determined  by Eq.(\ref{fraction-r}) assuming disk viscous parameter $\alpha=0.01$, $\beta=1$ for a $m=10^8$ black hole at the most efficient radiation region $r=49/12$. We show that the spectrum becomes steeper/softer at high accretion rates. Consequently, the bolometric correction from the 2-10~keV increases with increasing accretion rate.


We point out that, in the high accretion rate, the radiation pressure dominated disk is thermally and viscously unstable, which can lead to the S-shaped ``limit-cycle'' (e.g. Shakura \& Sunyaev 1976; Abramowicz et al. 1988; Honma, Matsumoto \& Kato 1991; Szuszkiewicz \& Miller 1998; see however, Hirose et al.(2009) who claimed no instability by local radiation-hydrodynamical simulations). However, if a fraction of accretion energy is liberated in the corona or the outflow the disk can be stable (Nakamura \& Osaki 1993; Svensson \& Zdziarski 1994). The magnetic heating to the corona discussed here is helpful to stabilize the disk, though $f$ is not large enough.

\subsection{Effect of viscous heating directly to electrons}
In our corona, we assume that all of the viscously dissipated energy heats ions only. However, direct heating to the electrons can also increase the corona emission. According to the detailed modeling to Sgr A* (Yuan et al. 2003), a significant fraction of the viscously dissipated energy should heat electrons directly. Such a result was later supported by the numerical simulation done by Sharma et al. (2007). Nevertheless, this effect is negligible if much larger additional heating to the corona is assumed. When a strong disk exists under a corona as in luminous AGN, a small fraction of disk accretion energy transported to coronal electrons can dominate the coronal heating.  For instance, assuming $\dot m=0.2$ and $f=0.1$,  the additional heating to the corona is 0.02 (in unit of Eddington luminosity); While the accretion rate in the corona, as shown in Fig.4, is $10^{-4}$ to $10^{-3}$ in the major radiation region, which produces viscous heat of $10^{-4}$ to $10^{-3}$ (in unit of Eddington luminosity). Even if all of this viscous heating directly goes to electrons, it is much smaller than the additional heating 0.02. The ratio is in the range of  0.5\%  to  5\% . This ratio is nearly the same for a higher $f$ since the accretion rate in the corona increases with increasing $f$.  For an accretion rate higher than 0.2, the ratio is smaller, which can be derived from the accretion rate of corona shown in the right panel of Fig.4.  Therefore, the neglect of viscous heating directly to electrons is a reasonable approximation.

\subsection{Disk Wind at High Accretion Rate}
When the accretion is greater than $0.1\dot{M}_{\rm Edd}$, the line driven wind and radiation  pressure driven wind will be important, which results in the accretion rate decreasing with the decrease of radius as $\dot{M}\propto R^{a}$ ($0.0<a<1.0$)(see Proga et al. 2000; Ohsuga et al. 2005). It is clear that with wind loss  the soft-photon field becomes weaker than the case without winds. Therefore, the Compton cooling is weaker, leading to less condensation. When the system reaches equilibrium, the spectrum can be harder.  This is the case without additional energy input. In this work, an additional energy is added into the corona and we find that the larger the $f$ is, the harder the spectrum is.  Since $0<a<1$, the energy generating rate $\propto \frac{GM \dot{M}}{R} \propto \frac{\dot{M}_{\rm out}}{R^{1-a}}$ is still dominated by  the innermost region. Thus, the wind loss  causes a decrease of disk luminosity and hence a harder spectrum.  However, if wind mass loss is very significant, there could be no longer optically thick disk and the effects are large.

\subsection{Irradiation}
A corona  lying above a cool disk can illuminate the disk, which increases the energy density of soft photons, leading to stronger coronal radiation.  This effect can be significant in the case of a strong corona above a truncated disk. However, for a relatively weak corona above a strong disk, as the case of HLAGNs, the effect of irradiation is unimportant. We estimate this effect  as follows.

Photons emitted from corona are roughly isotropic, about half of which go down to a slab-like disk,  irradiating the disk. The irradiating flux is,

\begin{equation}\label{e:irradiate1}
F_{\rm irr}=\frac{1}{2}F_{c}(1-a)=\frac{1}{2}f F_{\rm tot}(1-a),
\end{equation}
 where $a$ is  albedo which is usually taken as 0.15 (e.g. Zdziarski et al. 1999).
 This flux is scattered/absorbed in the disk and  eventually re-emitted, contributing to the soft photon field.  The ratio of this illuminating flux to the disk radiation, $F_{\rm d}=F_{\rm tot}(1-f)$,  is
 \begin{equation}\label{e:irradiate1}
{F_{\rm irr}\over F_{\rm d}}={\frac{1}{2}f(1-a)\over (1-f)}.
\end{equation}
This ratio reaches 1 when $f > 0.7$, indicating that the irradiation becomes dominant only when a large fraction of accretion energy is released in the corona.  Therefore, we don't include the irradiation in this work in order to make the numerical calculations simple.

\subsection{Comparison with the MHD Results}
 Hirose, Krolik \& Stone (2006) performed the shearing box numerical simulation with radiative transport, focusing on the vertical structure of the standard thin disk around $300R_{\rm g}$ ($R_{\rm g}=\frac{GM}{c^2}$) . They found
 that there is actually very little dissipation in  the coronal region as the corona is magnetically dominated. Consequently, the corona is too weak to explain the hard X-ray emission observed in AGNs. The disk corona evaporation/condensation model without additional heating gives similar results as MHD simulation (for details see Meyer-Hofmeister, Liu, \& Meyer 2012). When a corona is heated only by viscous dissipation within the corona, efficient inverse Compton scattering of the disk photons leads to over-cooling of the corona. As a consequence, part of the coronal gas condenses into the disk, leaving a very weak corona.

     To prevent the corona from collapsing, we assume a fraction of accretion energy is transported from the disk and release into corona in this paper, which appears conflict with the MHD simulation results. We note that in MHD simulation of Hirose et al. (2006), it was presumed that the disk is dominated by gas pressure, which is reasonable for studying the vertical structure of black hole X-ray binaries in about $300R_{\rm g}$. However, we are investigating the inner region of AGN disk, where the disk is radiation-pressure dominant. Recently, Blaes et al.(2011) simulated such kind of accretion flow and found that the thermodynamics of a radiation-pressure dominant  disk differs significantly from that envisaged in standard static models of accretion disks. In radiation-dominant plasma, the buoyant motions become significant for the overall energetics of the plasma. The photons' outward advection becomes comparable to radiative diffusion, and the associated vertical expansion work balances vertical heat transport and dissipation. At the present stage, it is not clear how much discrepancy exists between MHD simulation and our results.

\subsection{Luminous Hot Accretion Flow as an Alternative Model}
Based on the thermal stability analysis presented in Yuan (2003), luminous hot accretion flow (LHAFs, Yuan 2001) could also have a two-phase structure. Yuan \& Zdziarski (2004) calculated the emitted luminosity from such kind of two-phase accretion flow and found that it can produce X-ray luminosity as high as $10\%$ Eddington luminosity.  This is an alternative model for AGNs with intermediate spectrum and Eddington ratio. In our work,  we aim at more general case for HLAGNs with luminosity up to Eddington value, and calculate the broad waveband spectra from  both the disk and corona.

\section{Conclusion}

We applied the disk evaporation model to the high-luminosity AGNs.
To explain the typical spectra of HLAGNs, we conclude that there should be additional heating to the corona to prevent it from over-cooling by strong inverse Compton scattering.

  We assume that a fraction of gravitational
  energy ($f$) is liberated in the corona and calculate
  the corona structure. Then the  spectrum from the disk and corona is calculated by Monte Carlo simulation.
  We find that the hard X-ray is
  dominated by Compton cooling and the photon index  for hard X-ray in $2-10\rm~keV$ is $2.2<\Gamma <2.7$.
  For a given accretion rate, $\Gamma$ and $L_{bol}/L_{\rm 2-10 kev}$ decrease with increase of $f$.
   We discuss a possible mechanism for the presumed heating to the corona, that is, the magnetic heating. With equipartition of magnetic energy to the gas energy in the disk, we derive the fraction of accretion energy stored in magnetic field. Assuming this energy is released in the corona,  the  model predicts that, for $f<\frac{8}{35}$, the hard X-ray becomes softer at higher accretion rate and $L_{bol}/L_{\rm 2-10 kev}$
  increases, which is roughly consistent with the observational results.

\textbf{Acknowledgements} We thank the referee's valuable comments and suggestions. This work is supported by the National
Natural Science Foundation of China (Grant Nos. 11033007 and 11173029) and the National Basic
Research Program of China-973 Program 2009CB824800.

{}

\end{document}